 \documentclass[final,5p,times,twocolumn,authoryear]{elsarticle}

\usepackage{amssymb}
\usepackage{lipsum}
\usepackage{aas_macros}
\usepackage{tikz}
\usetikzlibrary{shapes,arrows}

\usepackage[breaklinks,colorlinks=true,citecolor=black,linkcolor=black,urlcolor=blue,bookmarks=true]{hyperref}

\journal{Astronomy $\&$ Computing}
\begin{document}

\begin{frontmatter}



\title{Search for the edge-on galaxies using an artificial neural network}
\author[spbu,gao,sao]{S.S.\ Savchenko}
\author[sao]{D.I.\ Makarov}
\author[sao]{A.V.\ Antipova}
\author[mpe]{I.S.\ Tikhonenko}

\affiliation[spbu]{organization={Saint Petersburg State University, 
                                 Department of Astrophysics},
                  addressline={University Embankment, 7/9}, 
                  city={St. Petersburg},
                  postcode={199034}, 
                  country={Russia}}
\affiliation[gao]{organization={Pulkovo Observatory of the Russian Academy of Sciences},
                  addressline={Pulkovskoye Shosse, 65}, 
                  city={St. Petersburg},
                  postcode={196140}, 
                  country={Russia}}
\affiliation[sao]{organization={Special Astrophysical Observatory, Russian Academy of Sciences},
                  addressline={Nizhnii Arkhyz}, 
                  postcode={369167}, 
                  state={Karachai-Cherkessian Republic},
                  country={Russia}}
\affiliation[mpe]{organization={Max-Planck-Institut für extraterrestrische Physik},
                  addressline={Gießenbachstraße 1}, 
                  postcode={D-85748}, 
                  city={Garching},
                  country={Germany}}

\begin{abstract}
We present an application of an artificial neural network methodology to a modern wide-field sky survey Pan-STARRS1 in order to build a high-quality sample of disk galaxies visible in edge-on orientation.
Such galaxies play an important role in the study of the vertical distribution of stars, gas and dust, which is usually not available to study in other galaxies outside the Milky Way.
We give a detailed description of the network architecture and the learning process.
The method demonstrates good effectiveness with detection rate about 97\% and it works equally well for galaxies over a wide range of brightnesses and sizes, which resulted in a creation of a catalogue of edge-on galaxies with $10^5$ of objects. The catalogue is published on-line with an open access.
\end{abstract}

\end{frontmatter}

\section{Introduction}
\label{introduction}

\begin{figure}
\centering
\includegraphics[width=\columnwidth]{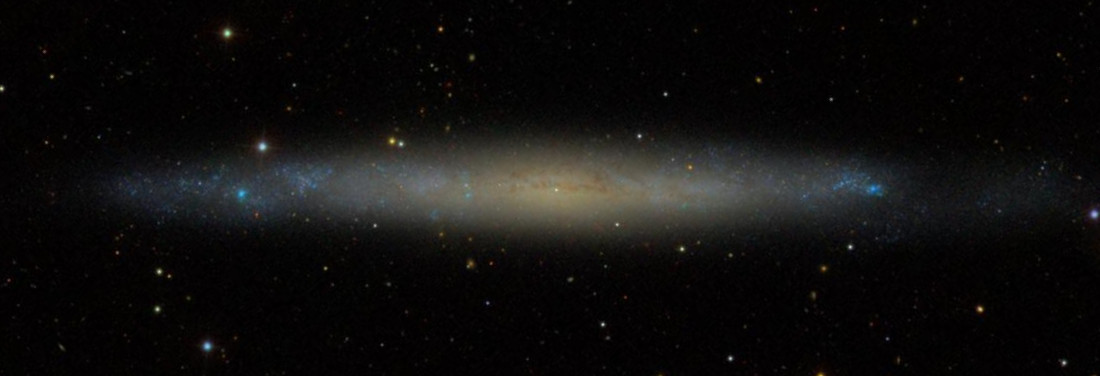}
\hspace{-3mm}\makebox[0pt][r]{\raisebox{3mm}{\textcolor{white}{NGC 4244}}}\hspace{3mm}
\includegraphics[width=\columnwidth]{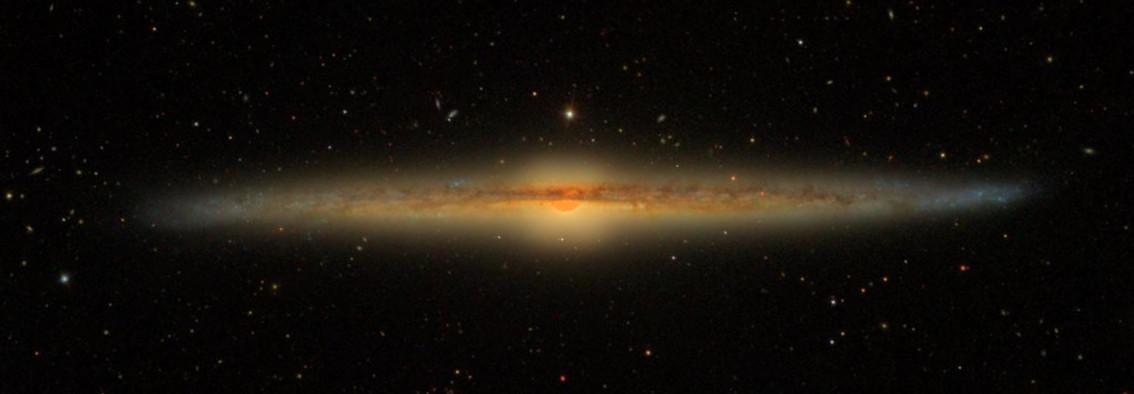}
\hspace{-3mm}\makebox[0pt][r]{\raisebox{3mm}{\textcolor{white}{NGC 4565}}}\hspace{3mm}
\includegraphics[width=\columnwidth]{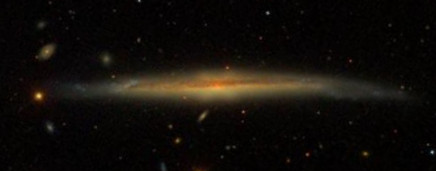}
\hspace{-3mm}\makebox[0pt][r]{\raisebox{3mm}{\textcolor{white}{NGC 5529}}}\hspace{3mm}
\caption{
Examples of edge-on galaxies as seen in SDSS images. 
Thanks to the orientation, we clearly see their substructures such as the dust lane in all three galaxies, boxy/peanut bulge of NGC~4565 and X-structure of NGC~5529, as well as a disk warp of NGC~5529.
Otherwise, it would be extremely difficult to detect these features.
}
\label{fig:Edge-on}
\end{figure}

Galaxies oriented almost edge-on to the observer form an important subset of all disk galaxies, since only these extragalactic objects allow us to directly observe the vertical distribution of stars, gas and dust. 
Some disk structures like boxy/peanut shaped bulges~\citep{Burbidge1959}, chimneys~\citep{Reach2020}, halos~\citep{Mosenkov2020}, and some properties of disks themselves like flaring and warping appear in all their glory only in edge-on galaxies.
Even if some galactic component is not a part of a disk, as in the case of the polar rings~\citep{Whitmore1990} and the polar bulges~\citep{Reshetnikov2015}, it is still best seen in a highly inclined galaxy due to the projection effects. Examples of edge-on galaxies obtained in the Sloan Digital Sky Survey are shown in Fig.~\ref{fig:Edge-on}. 

Within a framework of a comprehensive study of edge-on galaxies, it is necessary to create a new representative sample of such galaxies. 
A sample should be based on a wide-field survey to ensure uniformity of imaging data across the sky, include as many objects as possible in an orientation as close as possible to the edge-on orientation, and cover all Hubble types of disk galaxies.

\begin{figure*}
    \centering
    \includegraphics[width=0.3\textwidth]{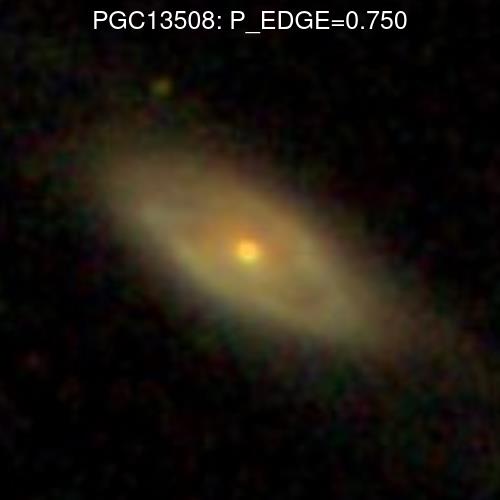}
    \includegraphics[width=0.3\textwidth]{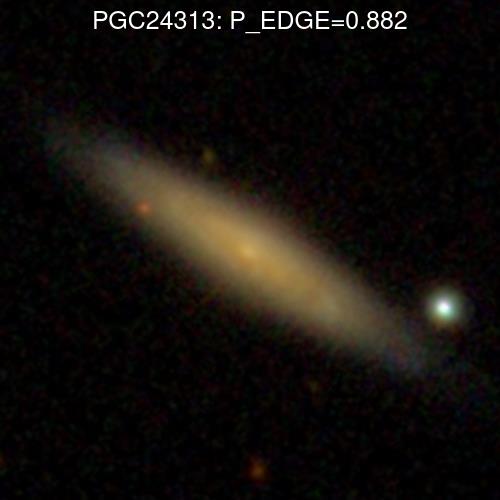}
    \includegraphics[width=0.3\textwidth]{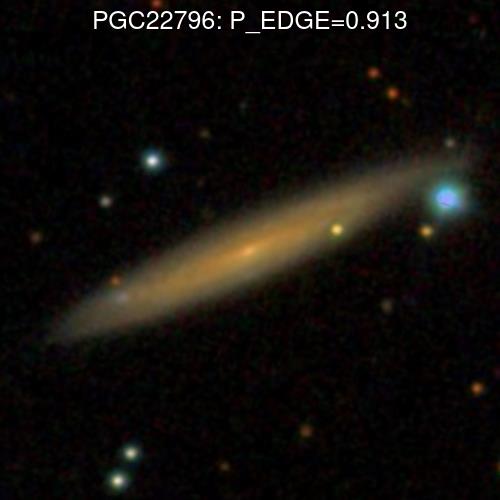}
    \caption{Examples of galaxies from the Galaxy Zoo~1 sample with their P\_EDGE values.}
    \label{fig:galaxy_zoo}
\end{figure*}

\begin{figure}[h]
\centering
\includegraphics[width=\columnwidth]{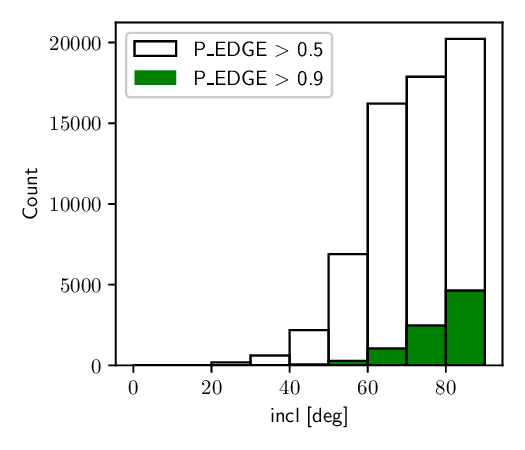}
\caption{A distributions of galaxies from Galaxy Zoo~1 catalogue by their inclinations according to HyperLEDA database.  The white histogram: galaxies with $\mbox{P\_EDGE} > 0.5$, the green one -- $\mbox{P\_EDGE} >0.9$.}
\label{fig:leda_inclinations}
\end{figure}

At the moment there were not many catalogs that satisfy this condition.
For example, the Revised Flat Galaxy Catalog \citep[RFGC,][]{Karachentsev1999} 
was formed by a visual search for galaxies on photographic prints of the Palomar Observatory (POSS-I) and the ESO/SERC sky surveys. It contains 4236 thin, $a/b\ge7$, mainly late-type spirals. 
Another example is the catalog of genuine edge-on disk galaxies \citep[EGIS,][]{Bizyaev2014} that has been prepared 
on a base of the Seventh Data Release of the Sloan Digital Sky Survey \citep[SDSS,][]{SDSSdr7}. In about one third of the sky, 5,747 edge-on galaxies were detected without significant morphological discrimination. 
A larger list of edge-on galaxies can be obtained via a crowdsource classification of a big sample of galaxies. For example, the  Galaxy Zoo~1 catalog \citep{Lintott2008} contains such a classification for 667,945 galaxies. For each object, the percentage of votes given for it being an edge=on galaxy is noted. The downside of this approach is that the classification was made by non-specialists, and therefore substantial bias and misclassifications are possible. 
Moreover, the strictness of the concept of ``the edge-on galaxy'' can vary very significantly between volunteers, and as a result, moderately inclined objects can be classified as the edge-on galaxies.
According to HyperLEDA\footnote{\url{http://leda.univ-lyon1.fr/}} database \citep{Makarov2014} the median inclination of 70,478 galaxies from the Galaxy Zoo~1 sample with vote fraction $\mathrm{P\_EDGE}>0.5$ is $72^\circ$. 
Figure~\ref{fig:galaxy_zoo} shows examples of galaxies that have high P\_EDGE values, but they fit poorly the concept of ``the edge-on orientation to the observer'':
in all three cases a disc structure is clearly visible. 
The situation improves, if one imposes a more strict limit on the P\_EDGE value, but this leads to significant decrease of the number of selected galaxies: there are only 9,300 galaxies with $\mbox{P\_EDGE} >0.9$. Fig.~\ref{fig:leda_inclinations} shows the distribution of galaxies from Galaxy Zoo~1 project by their inclinations according to HyperLEDA database for all galaxies with $\mbox{P\_EDGE} >0.5$ and $\mbox{P\_EDGE} >0.9$. One can see, that these samples include galaxies with orientations far from edge-on.

Our goal is to take advantage of the Panoramic Survey Telescope and Rapid Response System~\citep[Pan-STARRS,][]{Kaiser2010SPIE} survey to create a specialized catalog of edge-on galaxies that is bigger than the EGIS and RFGC, but cleaner than the one based on a crowdsource classification.
The second release of the survey is carried out in five broadband filters ($g$, $r$, $i$, $z$, $y$) with the first 1.8-meter diameter telescope, Pan-STARRS1, of the the Haleakala Observatory (Hawaii, US).
The Pan-STARRS1 survey covers three quarters of the sky above declination $\delta=-30^\circ$

The automatic detection of edge-on galaxies turns out to be a difficult task for a set of reasons. First of all, they appear in a great diversity of observed shapes: from bulge dominated red coloured Sa-type galaxies to almost bulgeless blue Sd ones. They can be smooth and almost featureless, but can be highly distorted and clumpy. Their disks can be perfectly flat, and can be extremely warped and asymmetrical. The second problem comes from the fact that the observed disk axis ratio does not unambiguously translate into its inclination: a thick but perfectly edge-on galaxy can have a lower axis ratio than a thin, but somewhat inclined one. 
Therefore, one can not rely solely on this parameter to separate edge-on galaxies from a general population: a lower value results in contamination of the sample by not perfectly edge-on galaxies, while higher one skews the sample to late type galaxies. 
Some other difficulties, such as contamination by foreground objects (mostly bright stars), the existence of a dust lane leading to a fictitious splitting of the galaxy into several separate objects, the presence of image artefacts that mimic real galaxies, further complicate the task.

In this article, we describe our approach to finding and identifying edge-on galaxies based on the use of artificial neural networks (ANN). This allowed us to significantly improve the quality of the candidate selection, reduce the project execution time, and to form the largest sample of edge-on galaxies to date.

Machine learning algorithms proved to be efficient tools for  classifications of galaxies. There are numerous publications where neural networks were applied both for annotating of growing samples of galaxies and for searching for galaxies of some specific type. For example, \citet{Dieleman2015} trained a convolutional neural network using visual morphological classifications of 55,000 galaxies. The model predictions accuracy was more than 99\% for galaxies, where there was a consensus among human participants annotating the training data. 
\cite{Domingues2018} used a convolutional neural network to improve the results of visual classifications of 14,034 galaxies published by \citet{Nair2010} and 304,122 galaxies published by \citet{Willett2013MNRAS}. 
Similar work was made by \citet{Walmsley2022}, where an ensemble of Bayesian convolutional networks was trained on visual classifications of 314,000 galaxies. 
An example of an unsupervised classification is presented in \citet{Spindler2021}, where a variational autoencoder was used for galaxy clustering and generative purposes.

Some works are focused not on the general morphology classifications, but on identifying specific features of galaxies or searching for galaxies of some specific type. 
Thus, \citet{Abraham2018} used a neural network to detect bar structures in images of almost face-on galaxies. 
\citet{Sarkar2023MNRAS} trained a model to distinguish between grand design and flocculent spiral patterns. 
The works by \citet{Ackermann2018MNRAS} and \citet{Domingues2023} were dedicated to the detection of galaxy merging and tidal features. 
The work we present in this paper falls into this class. 
We are creating a model to search specifically for galaxies in an edge-on orientation.

\section{First attempt}

At a preliminary stage, we tried to extract edge-on galaxy candidates from a catalog of extended sources found by the Pan-STARRS1 data processing pipeline \citep{PS1SourceDetection}. 
This catalogue contains the results of fitting of all objects with Kron magnitudes smaller than $21.5$, $21.5$, $21.5$, $20.5$, $19.5$~mag in $g$, $r$, $i$, $z$, $y$ bands respectively by the S\'ersic profile~\citep{Sersic1963}.  
Our search was carried out only outside the Zone of Avoidance defined by $|b| > 20^\circ + 15^\circ \exp(-\frac{1}{2}(\frac{\ell}{50^\circ})^2)$. 

\begin{figure}
\centering
\includegraphics[width=0.49\linewidth,trim=0cm 0cm 1.1cm 1cm,clip]{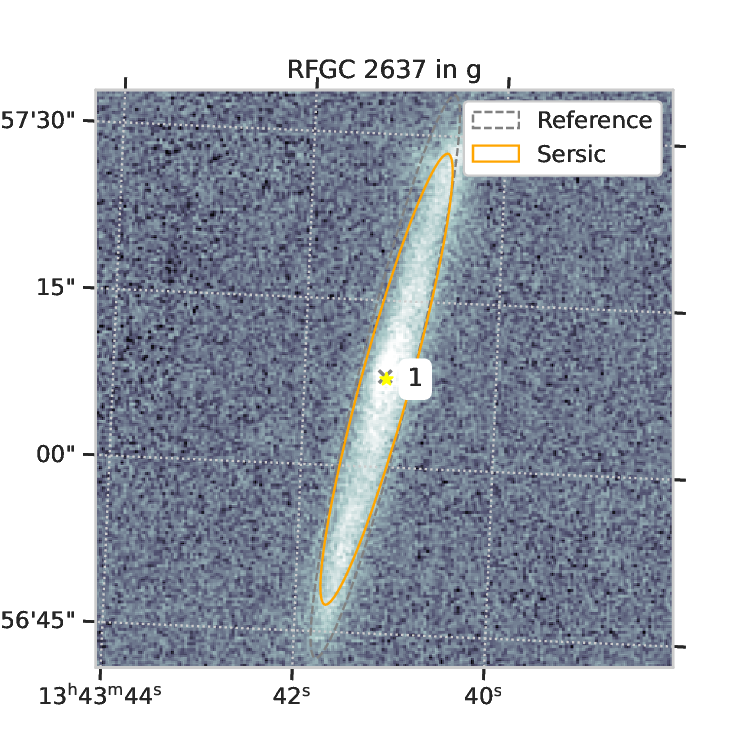}\hfill
\includegraphics[width=0.49\linewidth,trim=0cm 0cm 1.1cm 1cm,clip]{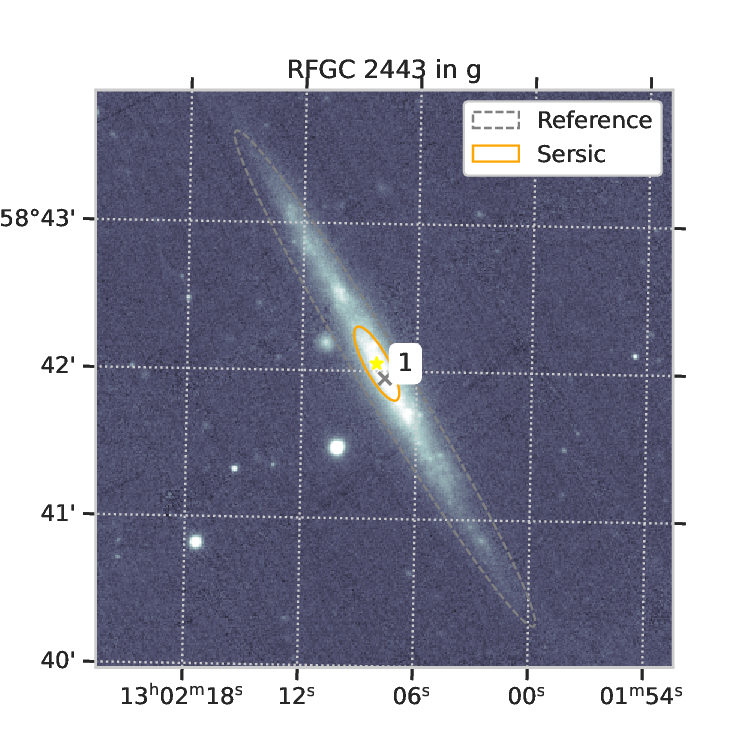}
\caption{
The left image illustrate a good match between size estimates in Pan-STARRS and RFGC catalogs,
while the case of the disagreement is shown in the right image.
Grey dashed ellipses, labelled `Reference' indicate the galaxy contour from RFGC, 
while orange ellipses are drawn from the parameters of a Sersic fit in PanSTARRS data (\texttt{gSerMajor}, \texttt{gSerAb}).
}
\label{fig:galaxypsrfgcexamples}
\end{figure}

To develop a selection criterion for the edge-on galaxies, we cross-matched flat galaxies from the RFGC catalog \citep{Karachentsev1999} with Pan-STARRS1 extended sources in their common domain in the sky.
Surprisingly, we did not find clear correlations between the diameters of RFGC galaxies and the half-light radii of automatically selected objects in Pan-STARRS survey.
The correlation between the axes ratios was not found as well.
This discrepancy rises from the difference in the source selection algorithms employed in the catalogs.
The sizes of the RFGC-galaxies have been measured by experts and closely follow the apparent shape of the galaxies, while the Pan-STARRS1 pipeline focuses on compact objects and often selects only the brightest central region of a galaxy or splits a galaxy into multiple pieces.
In both cases, this leads to an incorrect size estimate.
The resulting candidates tend to be smaller and rounder then their RFGC counterparts (see Fig.~\ref{fig:galaxypsrfgcexamples} for the cases of a good match and the aforementioned problem).

As a result, we concluded that the desired catalog of edge-on galaxies can not be easily constructed by constraining object parameters from the Pan-STARRS1 database, and a more advanced technique should be used instead.

\section{Deep learning for images classification}
\label{sec:nn}

\begin{figure*}
  \centering
  \tikzstyle{block} = [rectangle, draw, fill=blue!15,
    text width=5.5em, text centered, rounded corners, minimum height=2em]
\footnotesize
\begin{tikzpicture}[node distance = 2.5cm, auto]
  % Place nodes
  \node [block] (input) {Input $48 \times 48 \times 3$};
  \node [inner sep=0pt] (imageg) at (0, -2.5) {\fcolorbox{black}{white}{\includegraphics[width=.125\textwidth]{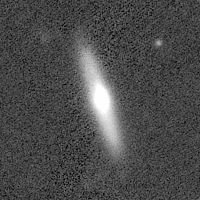}}};
  \node [inner sep=0pt] (imager) at (0.25, -2.75) {\fcolorbox{black}{white}{\includegraphics[width=.125\textwidth]{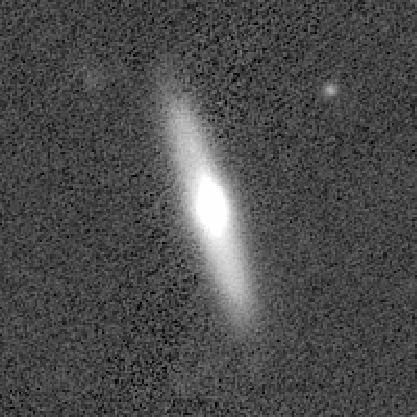}}};
  \node [inner sep=0pt] (imagei) at (0.5, -3) {\fcolorbox{black}{white}{\includegraphics[width=.125\textwidth]{images/cand_g}}};
  \node [text width=3cm, below of=imagei, node distance=1.8cm, align=center] {Original images in g, r and i bands};
  \node [block, right of=input, node distance=3cm] (conv11) {Conv $(5 \times 5)$ $16\times 3$};
  \node [block, below of=conv11, node distance=1.3cm] (conv12) {Conv $(5 \times 5)$ $16\times 3$};
  \node [block, below of=conv12, node distance=1.3cm] (batch1) {Batch norm};
  \node [block, below of=batch1, node distance=1.1cm] (pool1) {Max-pool $2\times 2$};
  \node [block, below of=pool1, node distance=1.1cm] (drop1) {Dropout 0.3};
  \node [block, right of=conv11, node distance=2.5cm] (conv21) {Conv $(3 \times 3)$ $32\times 3$};
  \node [block, below of=conv21, node distance=1.3cm] (conv22) {Conv $(3 \times 3)$ $32\times 3$};
  \node [block, below of=conv22, node distance=1.3cm] (batch2) {Batch norm};
  \node [block, below of=batch2, node distance=1.1cm] (pool2) {Max-pool $2\times 2$};
  \node [block, below of=pool2, node distance=1.1cm] (drop2) {Dropout 0.3};
  \node [block, right of=conv21, node distance=2.5cm] (conv31) {Conv $(3 \times 3)$ $64\times 3$};
  \node [block, below of=conv31, node distance=1.3cm] (conv32) {Conv $(3 \times 3)$ $64\times 3$};
  \node [block, below of=conv32, node distance=1.3cm] (batch3) {Batch norm};
  \node [block, below of=batch3, node distance=1.1cm] (pool3) {Max-pool $2\times 2$};
  \node [block, below of=pool3, node distance=1.1cm] (drop3) {Dropout 0.3};
  \node [block, right of=batch3, node distance=2cm, minimum height=5.5cm, text width=1.5em] (fc1) {\rotatebox{-90}{Full 500}};
  \node [block, right of=fc1, node distance=1.5cm, minimum height=2.5cm, text width=1.5em] (fc2) {\rotatebox{-90}{Full 2}};
  \node [block, right of=fc2, node distance=2cm] (out) {Object type};

  % % Draw edges
  \draw [-, black, -latex'] (imageg) -> node[right, text width=2.25cm]{Downscaling+ normalisation} (input);
  \draw [-, black, -latex'] (input) -> (conv11);
  \draw [-, black, -latex'] (conv11) -> node[right]{ReLU} (conv12);
  \draw [-, black, -latex'] (conv12) -> node[right]{ReLU} (batch1);
  \draw [-, black, -latex'] (batch1) -> (pool1);
  \draw [-, black, -latex'] (pool1) -> (drop1);
  \draw [-, black, -latex'] (drop1) -- ++(1.25cm,0) |-  (conv21);
  \draw [-, black, -latex'] (conv21) -> node[right]{ReLU} (conv22);
  \draw [-, black, -latex'] (conv22) -> node[right]{ReLU} (batch2);
  \draw [-, black, -latex'] (batch2) -> (pool2);
  \draw [-, black, -latex'] (pool2) -> (drop2);
  \draw [-, black, -latex'] (drop2) -- ++(1.25cm,0) |-  (conv31);
  \draw [-, black, -latex'] (conv31) -> node[right]{ReLU} (conv32);
  \draw [-, black, -latex'] (conv32) -> node[right]{ReLU} (batch3);
  \draw [-, black, -latex'] (batch3) -> (pool3);
  \draw [-, black, -latex'] (pool3) -> (drop3);
  \draw [-, black, -latex'] (drop3) -- ++(1.25cm,0) |-  (fc1);
  \draw [-, black, -latex'] (fc1) -> node[below]{\rotatebox{-90}{ReLU}} (fc2);
  \draw [-, black, -latex'] (fc2) -> node[below]{\rotatebox{-90}{SoftMax}} (out);
\end{tikzpicture}
\normalsize
  \caption{Convolutional neural network architecture. The neural network consists of three rounds of double convolutional
  blocks with batch normalisation, max-pooling and dropout layers each, and two fully connected layers at the end. For convolutional
  blocks, the first line shows the convolution size, the second one -- the number of convolutions in each block (times three for three
  passbands). All max-pooling layers have $2\times2$ size and all dropout layers have drop fraction equal to 0.3.}
  \label{fig:net_arch}
\end{figure*}

In this work we use an artificial neural network (ANN) algorithm to perform the classification of the images. 
The network consists of interconnected \textit{neurons}, each with multiple inputs and an output that mimic the synapse of a biological brain, and in its simplest form, it can be mathematically described as a weighted sum:
$$
y = f\left(\sum_{j=1}^n w_i x_i + w_0\right),
$$
where $x_i$ are inputs and $w_i$ are so called weights, and $f(x)$ is an activation function, that adds non-linearity to the model. 
The neurons of a network are organised in \textit{layers} such that the outputs of one layer neurons are connected to the inputs of the next layer neurons. 
The first layer of neurons is the input layer and its inputs $x_i$ are the data values. 
As it propagates through the network, the data is modified at each layer according to the weights until it reaches the last layer from which it can be read.
In other words, the network maps the input vector of the data to some output vector.

The result of this data modification depends on the architecture of the neural network (the number of neurons, the number of layers and their structure) and the weights values. 
In general, a more complex network allows more complex behaviour.
Common sense is to start with relatively simple architecture and build it up until the required level of complexity is reached.
Once the network architecture is fixed, the weights values can be tuned to achieve desired network behaviour. 
This can be done through supervised learning process. 
The key ingredient is a training dataset that consists of pairs of vectors $\vec{x}_i$ that represent the input data and $\vec{y}_i$ that represent the desired output for $\vec{x}_i$. 
For each training example the weights of the network are modified using some optimisation technique (such as the stochastic gradient descent algorithm) in a backpropagation manner: weights are modified starting from the last layer back to the first one. As a result, after each iteration the network becomes better in mapping of a given input vector $\vec{x}_i$ to a given output vector $\vec{y}_i$. 
The training process continues until the network converges to some steady state or some specified precision is reached.

One of the most widely used applications of ANNs is the classification problem. 
In this case, the input vector is the data to be classified and the output vector represents the class, and the network is trained to properly map the input data to the correct class.
In the simplest case of two target classes, the network output can be a scalar with different values for different classes (usually 0 for one class and 1 for another).
In the context of our task of creating a catalog of edge-on galaxies, a natural approach is to have two classes:
the class 1 (positive case) corresponds to an image of the edge-on galaxy, and class 0 (negative case) is an image of any other object.

% In this section we describe all about the ANN learning process: the training data, the network architecture, and the learning process.

\subsection{The network architecture}
\label{sec:networkArchitecture}

The architecture of a neural network is described by the number and type of layers, the number of neurons in each layer,
the size and number of a convolutional kernels, etc.
All these values together are called neural network hyperparameters. 
During the training process, the hyperparameters are fixed, and only the weights are changed. 
The best network architecture depends on a specific problem, and there is no easy way to build the optimal network for a given task.

The usual way to solve this is to use a trial and error approach: by modifying the hyperparameters of a neural
network and estimating its resulting performance, one can gradually build a better network until the desired quality is
achieved. 
The common practice is to start with a relatively simple network and move towards a more complex one keeping the track of the overall performance. 
It is worth to note that a wide variety of different architectures can provide the similar performance, but among them it is better to choose one with a simpler architecture (with lower number of tunable weights), as it is easier to train.

In this work, we use a convolutional neural network (CNN) to perform galaxy classification. 
A CNN is a specialized type of ANN designed to process data that has a spatial structure such as images.
CNN neurons are organised in spatial filters that perform discrete convolution of the image $F$ at coordinates $m$, $n$ with the convolution matrix $K$:
%, and result of an application of such filter $K$ to a segment of image $F$ at coordinates $m, n$ is a discreet convolution:
$$
G\left[m, n\right] = (F*K)\left[ m, n \right] = \sum_i \sum_j K\left[ i, j \right] F \left[ m-i, n-j \right],    
$$
where indices $i$ and $j$ go over all elements of the kernel $K$. 
By sliding the filter kernel over the entire input image, one can compute the full response image $G$, the size of which depends on the convolution parameters: the kernel size, the stride, and the padding~\citep{dumoulin2016guide}. 
The output image can be fed forward as an input to the next layer (convolutional or not) of the network. 
\citet{LeCun1989, LeCun1990} showed that including such convolutional layers into the network architecture greatly improves network performance in case of the image processing.

Even though a network architecture can be fully convolutional, i.e.\ consisting only of convolutional layers~\citep[for example, ][]{Long2015}, the image classification problem often involves a network in which several convolutional layers are followed by one or more fully connected layers. 
These fully connected layers map the output of the last convolutional layer into the output of the network.
In this approach, the convolutional part can be considered as a feature detector trained to search for certain patterns in the image, and the fully connected part is a classifier that uses these features to make a decision.

Figure~\ref{fig:net_arch} shows the final architecture of our network, which was adopted after the trial and error approach described above.

As an input, the network takes a stack of three images of an object with a size of $48 \times 48$ pixels in three $g$, $r$, $i$ Pan-STARRS1 passbands. 
Therefore, the input is a 3D array of size $48\times 48\times 3$. 

The network consists of three rounds of convolutional blocks. 
Each block has two convolutional layers followed by batch normalisation, max-pooling and dropout layers. 
The convolutional layers of the first block have 16 kernels of $5\times 5$ pixels for each of the three bands. 
The convolutional layers of the second and the third blocks have 32 and 64 kernels of $3 \times 3$ pixels for each band, respectively.
We use the rectified linear unit \citep[ReLU,][]{Hahnloser2000} as an activation function for the convolution layer  neurons.

In each block, the second convolutional layer is followed by batch normalisation, which normalizes the input using the mean and the standard deviation over a subsample of training examples (called a batch).
This reduces the variation in hidden layers during training and allows the network to converge faster~\citep{Ioffe2015}.

In each convolution block, the batch normalization is followed by a max-pooling layer.
During the pooling process, the data is split in patches of $2\times 2$ elements, and only maximal element of each patch is passed to the output. 
The pooling reduces the number of network elements to train, because only a quarter of neural connections are forwarded downstream, and also it makes the network less sensitive to small spatial translations of the input~\citep{Boureau2010}.

The last layer in the block is the dropout layer. 
It randomly removes a certain fraction of neurons along with all their connections on every pass during the training process~\citep{Srivastava2014}. 
Applying such modifications to the network reduces the problem of overfitting when the network remembers specific training examples rather than inferring general patterns from the training data. 
An overfitted network has high score on the training data, but does not generalize well to new data. 
We set the drop fraction to 0.3 in all three dropout layers of the network.

After the third round of convolution, there are two fully connected layers with 500 and 2 neurons, respectively, and the ReLU activation function. 
The output of the last layer is 
%the output of the entire network and 
the prediction of the type of the object, the image of which is fed to the input of the entire network.

Putting together all of the above, we get that the network contains 206,894 trainable parameters.

\subsection{Training dataset and data augmentation}
\label{sec:train_data}

To train a neural network as a classifier, one must provide it with a training dataset consisting of labelled data: a set of training examples with the correct class.
Even though we are only interested in edge-on galaxies (i.e.\ positive class), the training dataset must include both positive and negative examples, otherwise the network will converge to a trivial solution that always produces a positive output class.

As a source of positive examples, we used the EGIS catalog \citep{Bizyaev2014}, which includes 5,747 visually selected edge-on galaxies.
The vast majority of objects in this catalog are genuine edge-on galaxies that can be directly used as positive examples during the training process.

We used the Galaxy Zoo~1 catalog \citep{Lintott2008} to build a list of negative examples of non-edge-on galaxies.
The Galaxy Zoo~1 catalog contains rough visual classification of a large number of galaxies as fraction of votes made for different morphological classes (elliptical galaxy, spiral galaxy, edge-on galaxy, etc.). 
We selected all objects with a classification fraction for a edge-on galaxy (P\_EDGE) less than 0.1 (to exclude edge-on galaxies from the list of negative examples) and a vote fraction for an elliptical galaxy or a clockwise/anticlockwise spiral (P\_EL, P\_CW, P\_ACW) greater than 0.1. 
This step gave us a list of 270,000 galaxies. 

Although the crowdsource classification used in the Galaxy Zoo~1 project is subject to substantial bias, the main goal of this selection was achieved: visual inspection of a random subsample confirmed the absence of edge-on galaxies and the high diversity of all other galaxy types.

After cross-matching them with the HyperLEDA database,
we also imposed restrictions on the apparent diameter ($0.1<d_{25}<0.5$~arcmin) to exclude too small and unresolved objects, as well as too big and oversaturated galaxies, and on the apparent flattening ($q \ge 0.25$) to exclude disks close to edge-on orientation.
As a result, we formed a sample of negative examples for the ANN training, consisting of more than 54,000 galaxies that are non-edge-on disks.

Since most of the objects in the Pan-STARRS1 fields are stars, their images must also be included in the training sample as negative examples. 
For this purpose, we picked several dozens of stars from random Pan-STARRS1 fields.
Star images do not vary as much as galaxies do, therefore we do not need a great number of them to represent this class.

The final component of the negative examples is empty background fields. 
Spurious background variations can be detected as faint objects by image segmentation algorithm.
To take this into account, we extracted random regions of the images that do not contain any objects.

The resulting training dataset consists of 5,747 positive examples (edge-on galaxies) and over 54,000 negative examples (non-edge-on galaxies, stars and empty fields).

This dataset still faces two major problems that prevent successful network training. 
The first problem is its relatively small size.
Even though there are thousands of images, it is not enough to train a convolutional network that is deep enough to solve our classification problem.
A common solution in a situation where the sample size can not be easily increased by adding new objects is data augmentation.
The main idea is to apply some modifications to existing images to obtain new ones.
We applied several modifications:
\begin{itemize}
\item adding random Gaussian noise to the image;
\item random flipping of the image horizontally or/and vertically;
\item image rotation by a random angle;
\item image zoom by a random factor from 0.75 to 1.25;
\item shifting the image center randomly within 20\% of the image size.
\end{itemize}
Although this process creates new images that contain essentially the same objects, in practice, the sample augmented in this way leads to much better results in a model training, but the trade-off is that this method can introduce bias during the training process.

The second problem with the dataset is its skew in sample size (by about an order of magnitude) towards negative examples. 
Having significantly more negative examples during training, the network tends to converge to a solution that will always
predict a negative class (this would be true 9 times out of 10, so the formal testing score will be high even for this
trivial solution). 
To normalize the sample we adopted the following approach. 
Until the desired size of the training set is reached, we take one random image of an edge-on galaxy as a positive example and one random image of a negative one (with 80\% probability we take non-edge-on galaxy, 10\% a star and 10\% an empty field), apply augmentation algorithm to them and add them to the training set. 
At the end, the training sample will have exactly the same number of positive and negative examples, and its size can be arbitrary large, so both problems are solved. 

Our tests showed that the training quality increased after applying described above technique to our sample.

\subsection{Training process}
\label{sec:net_train}

%To train the network we have created a dataset using the described in Sec.~\ref{sec:train_data} data augmentation method. 
The dataset size described above was set to $3\cdot10^{5}$ objects. 
We split this entire dataset into two non-overlapping groups called the training dataset ($2\cdot10^{5}$ objects) and the test one ($10^5$ objects). 
The first is used to provide examples during the network training, while the latter is only used to evaluate the performance of the network. 
This approach of using two independent datasets is extermely important, because the network can overfit and demonstrate high performance score on the training data, but will not perform well on new data that was not used during the training. 
Computing the performance score on an independent dataset provides a less biased estimation of the network performance.

\begin{figure}
  \centering
  \includegraphics[width=\columnwidth]{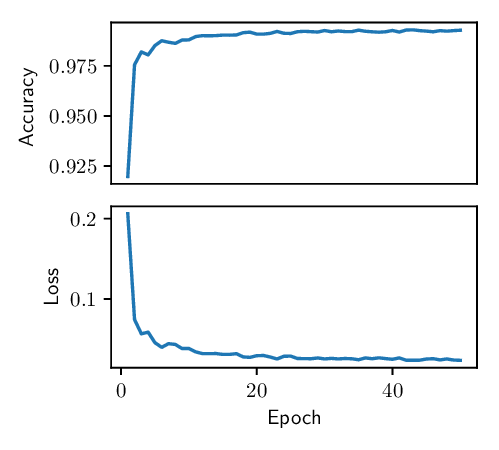}
  \caption{The network training history: the accuracy (top) and loss function value (bottom) as a function of epoch number.}
  \label{fig:train_history}
\end{figure}

The network was trained for 50 epochs (i.e.\ the training dataset was fed to the network 50 times) with categorical cross-entropy as the loss function. 
Fig.~\ref{fig:train_history} shows the network performance computed on the test dataset as a function of the epoch number. 
The upper panel of the figure shows the prediction accuracy, and the bottom one presents the loss function value. 
It is clearly seen that the accuracy of the network climbs rapidly to the level of 99\% during the first ten epochs, and after 20--30 epochs it reaches approximately equilibrium state.

For comparison, before applying data augmentation, the accuracy saturated at $\approx 95\%$ level.

Although $\approx 99\%$ accuracy seems rather good at first glance, it is not good enough for our goal of the edge-on galaxies search in the Pan-STARRS1 survey. 
The problem is the small fraction of edge-on galaxies among all the other objects on the celestial sphere: 
not all galaxies are disk ones, only a small fraction of disk galaxies are visible edge-on, and many objects in the processed images are not even galaxies (stars, asterisms, image artefacts). 
It turned out that among the two hundred selected objects, on average there is only one edge-on galaxy. 
This means that a trivial classifier that always classify objects as not edge-on galaxies would have an accuracy of 99.5\% (so called ZeroR classification accuracy). 
Of course, our trained model greatly surpasses this result, because its 99\% accuracy is achieved on a balanced training sample, for which the ZeroR classifier would only have accuracy of 50\%. 
But if applied to the real data as it is, our model would result in more false positive detections that there are real edge-on galaxies.

It is a challenge to improve the network accuracy above 99\%, as it faces many problems with limited number of training examples and the data overfitting is possible.
Instead, we used a different approach, sometimes called an ensemble classifier.
We created five independent training datasets using data augmentation and trained five networks that have the same architecture on them.

Even though all five networks have the identical architecture, they produce different predictions when trained on large augmented datasets.
We then use these five predictions on the object type and take a simple majority of the votes as the final result to reduce the chance of misclassification.

Thus, to get a false positive or false negative final classification, three out of five networks must make an erroneous prediction. This approach significantly improved the quality of classification.

\begin{table}
\centering
\caption{
The confusion matrix for our ensamble of 5 networks computed for 3027 edge-on galaxies from RFGC and the same amount of non-edge-on galaxies. 
Abbreviations are for true positive (TP), false negative (FN), false positive (FP) and true negative (TN).
}
\begin{tabular}{lll}
    & \multicolumn{2}{c}{Predicted} \\
    &        edge-on   & non-edge-on\\
\hline
Actual edge-on & $\mathrm{TP}=3010$ & $\mathrm{FN}=17$  \\
Actual non-edge-on & $\mathrm{FP}=25$ & $\mathrm{TN}=3002$ 
\end{tabular}
\label{tab:confusion}
\end{table}

As a final independent quality check, we applied our classification procedure to galaxies from the RFGC catalog~\citep{Karachentsev1999}. 
This catalog consists of 4236 thin edge-on galaxies and covers the entire sky. 
We used only those RFGC-galaxies that are not included into the EGIS catalog and, as a result, were not used for the network training. 
Among these 3027 unique RFGC galaxies, our network misclassified only 17 galaxies, giving the error rate less than 1\% (see Table~\ref{tab:confusion} for a full confusion matrix of the test).

\section{Search for edge-on galaxies}

\begin{figure*}
    \centering
    % Votes 0.5..0.6
    \includegraphics[width=0.09\textwidth]{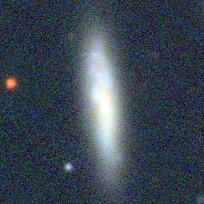}
    \includegraphics[width=0.09\textwidth]{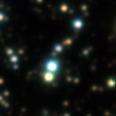}
    \includegraphics[width=0.09\textwidth]{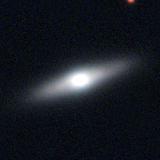}
    \includegraphics[width=0.09\textwidth]{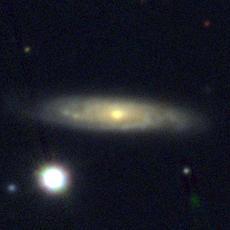}
    \includegraphics[width=0.09\textwidth]{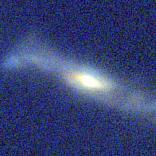}
    \includegraphics[width=0.09\textwidth]{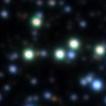}
    \includegraphics[width=0.09\textwidth]{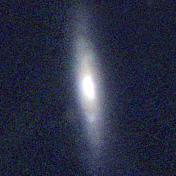}
    \includegraphics[width=0.09\textwidth]{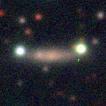}
    \includegraphics[width=0.09\textwidth]{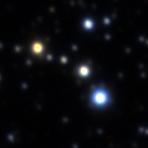}
    \includegraphics[width=0.09\textwidth]{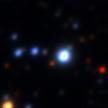}
    % Votes 0.6..0.7
    \includegraphics[width=0.09\textwidth]{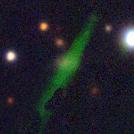}
    \includegraphics[width=0.09\textwidth]{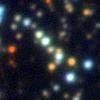}
    \includegraphics[width=0.09\textwidth]{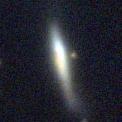}
    \includegraphics[width=0.09\textwidth]{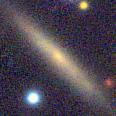}
    \includegraphics[width=0.09\textwidth]{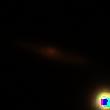}
    \includegraphics[width=0.09\textwidth]{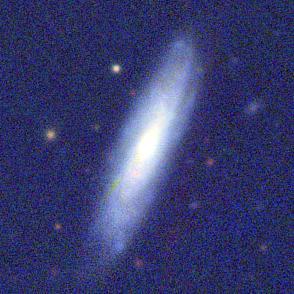}
    \includegraphics[width=0.09\textwidth]{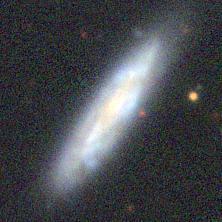}
    \includegraphics[width=0.09\textwidth]{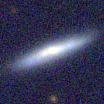}
    \includegraphics[width=0.09\textwidth]{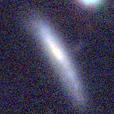}
    \includegraphics[width=0.09\textwidth]{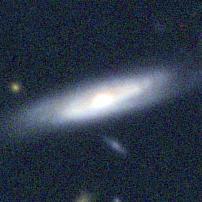}
    % Votes 0.7..0.8
    \includegraphics[width=0.09\textwidth]{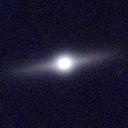}
    \includegraphics[width=0.09\textwidth]{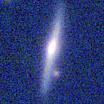}
    \includegraphics[width=0.09\textwidth]{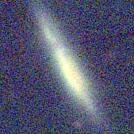}
    \includegraphics[width=0.09\textwidth]{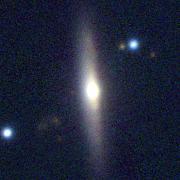}
    \includegraphics[width=0.09\textwidth]{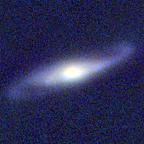}
    \includegraphics[width=0.09\textwidth]{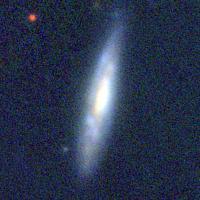}
    \includegraphics[width=0.09\textwidth]{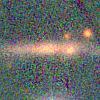}
    \includegraphics[width=0.09\textwidth]{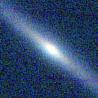}
    \includegraphics[width=0.09\textwidth]{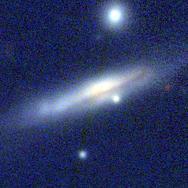}
    \includegraphics[width=0.09\textwidth]{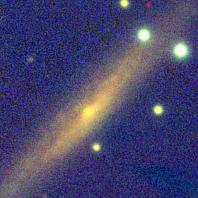}
    % Votes 0.8..0.9
    \includegraphics[width=0.09\textwidth]{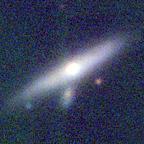}
    \includegraphics[width=0.09\textwidth]{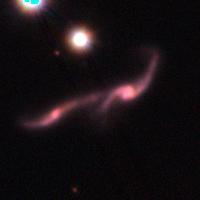}
    \includegraphics[width=0.09\textwidth]{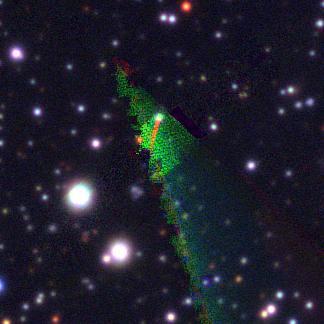}
    \includegraphics[width=0.09\textwidth]{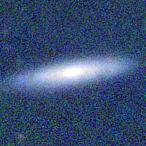}
    \includegraphics[width=0.09\textwidth]{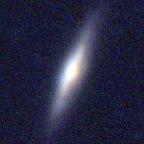}
    \includegraphics[width=0.09\textwidth]{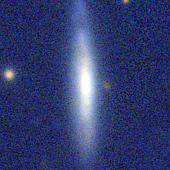}
    \includegraphics[width=0.09\textwidth]{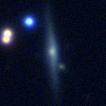}
    \includegraphics[width=0.09\textwidth]{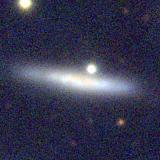}
    \includegraphics[width=0.09\textwidth]{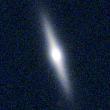}
    \includegraphics[width=0.09\textwidth]{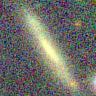}
    % Votes 0.9..1.0
    \includegraphics[width=0.09\textwidth]{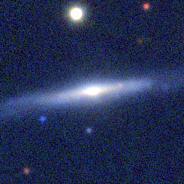}
    \includegraphics[width=0.09\textwidth]{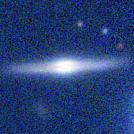}
    \includegraphics[width=0.09\textwidth]{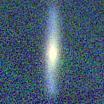}
    \includegraphics[width=0.09\textwidth]{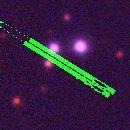}
    \includegraphics[width=0.09\textwidth]{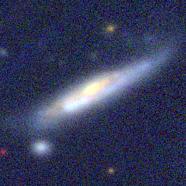}
    \includegraphics[width=0.09\textwidth]{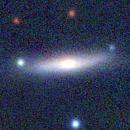}
    \includegraphics[width=0.09\textwidth]{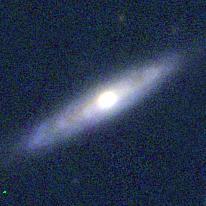}
    \includegraphics[width=0.09\textwidth]{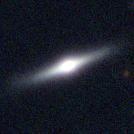}
    \includegraphics[width=0.09\textwidth]{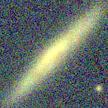}
    \includegraphics[width=0.09\textwidth]{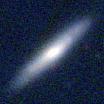}
\caption{
Examples of objects classified as edge-on galaxy by a single neural network from of the entire ensemble. 
Objects in different rows are grouped by the score assigned by the network, from the top row to the bottom:
$0.5 < \mathrm{score} \le 0.6$, $0.6 < \mathrm{score} \le 0.7$, $0.7 < \mathrm{score} \le 0.8$, $0.8 < \mathrm{score} \le 0.9$, and $0.9 < \mathrm{score} \le 1.0$.
}
\label{fig:1st_run_examples}
\end{figure*}

\subsection{Pipeline and the first run}
\label{sec:search}

The Pan-STARRS1 survey covers the celestial sphere north of declination $\delta=-30^\circ$. 
Access to public data~\citep{2020ApJS..251....7F} is provided by the Space Telescope Science Institute (STScI).
The survey image layout\footnote{\url{https://outerspace.stsci.edu/display/PANSTARRS/PS1+Sky+tessellation+patterns}} is organised as a regular grid of 2009 so-called projection cells.
The projection cell is an image of $63k\times63k$ pixels.
The pixel size is $0.25^{\prime\prime}$.
For convenience, each $4^\circ\times4^\circ$ projection cell is split in turn into a $10 \times 10$ grid of so-called skycells, each $24^\prime\times24^\prime$.
Taking into account that we are working in three photometric passbands, namely $g$, $r$ and $i$, we need to process about 600,000 skycell images.
Each skycell is a FITS-file~\citep{1981A&AS...44..363W} about 65~Mb each, so the total amount of data to process is about 5~Tb.

The search pipeline was organized as follows.
For each skycell, we download its images in $g$, $r$, and $i$ filters.
Next, we use the \textsc{detect\_sources} function of the \textsc{Python} package \textsc{photutils}~\citep{2020zndo...4044744B} to detect all objects in the $r$-band image. 
This segmentation algorithm takes two main parameters that govern the extraction of objects: the detection threshold level and the number of connected pixels with flux values above a given threshold. 
In our study, we choose the background variation threshold of $4\sigma$, and the number of connected pixels brighter than this threshold to be at least 15. 
Smaller values of these parameters lead to a large number of spurious detections due to random background variations. 
Further, we filter out all objects whose major axis is less than 48 pixels (the length of the detected object may be bigger than the number of connected pixels, because in the outer regions, some pixels may be disconnected from the main body of the object, but still be attributed to the object).
The size of 48 pixels, corresponding to $12^{\prime\prime}$, was chosen to provide a reliable thickness measurement for even thinnest galaxies, taking into account a typical seeing of $1.19^{\prime\prime}$ in $r$-band.
After that, we extract the images of each object; scale them to $48\times48$ pixels size according to the size of the input layer of the network; stack $g$, $r$, and $i$ images into a 3D array; and feed the array to our neural network quintet.
Finally, if at least three out of five networks score above 0.5 for the edge-on class, we mark this object as the edge-on galaxy and add it to our catalog. The downloaded field images can be deleted at this stage, so that the whole pipeline goes 'on the fly', without storing of the entire survey database locally.

To speed up the process, we run the search simultaneously on a dozen computers. 
One of them is used to keep track of processed skycells, to distribute jobs to others in a form of skycells IDs, and to collect the results. 
%All the other were used to perform computations. The whole survey was processed in 30 days.

\subsection{Visual inspection}
\label{sec:visualInspection}

When the search was complete, our pipeline returned a list of 26,719 objects. 
Figure~\ref{fig:1st_run_examples} shows examples of the found objects grouped by the scores assigned to them by the single network (i.e.\ without averaging over the whole ensemble of networks).
The top row shows objects with $0.5 < \mathrm{score} \le 0.6$, 
the objects on the second one have $0.6 < \mathrm{score} \le 0.7$, 
etc.\ down to the bottom row with objects $0.9 < \mathrm{score} \le 1.0$. 
It can be seen that the score clearly correlates with the quality of the selection. 
For low scores ($\approx 0.5$--0.7), there are many false detections of completely wrong objects, and even if an object is a disk galaxy, its orientation can be far from the edge-on. 
For higher scores, the fraction of genuine edge-on galaxies increases.

\begin{figure}
    \centering
    % Votes 0.5..0.6
    \includegraphics[width=0.09\textwidth]{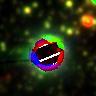}
    \includegraphics[width=0.09\textwidth]{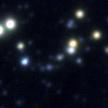}
    \includegraphics[width=0.09\textwidth]{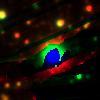}
    \includegraphics[width=0.09\textwidth]{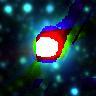}
    \includegraphics[width=0.09\textwidth]{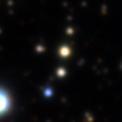}
    % Votes 0.6..0.7
    \includegraphics[width=0.09\textwidth]{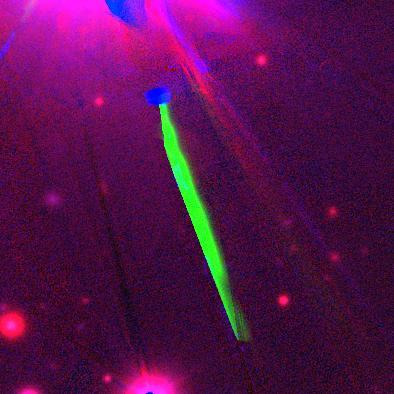}
    \includegraphics[width=0.09\textwidth]{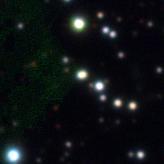}
    \includegraphics[width=0.09\textwidth]{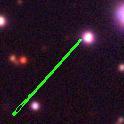}
    \includegraphics[width=0.09\textwidth]{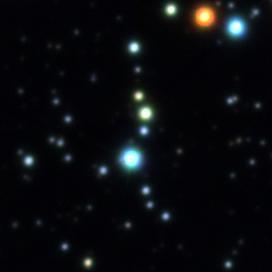}
    \includegraphics[width=0.09\textwidth]{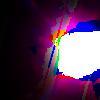}
    % Votes 0.7..0.8
    \includegraphics[width=0.09\textwidth]{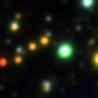}
    \includegraphics[width=0.09\textwidth]{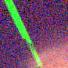}
    \includegraphics[width=0.09\textwidth]{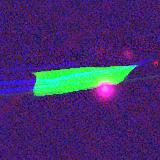}
    \includegraphics[width=0.09\textwidth]{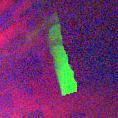}
    \includegraphics[width=0.09\textwidth]{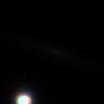}
    % Votes 0.8..0.9
    \includegraphics[width=0.09\textwidth]{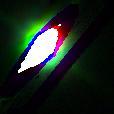}
    \includegraphics[width=0.09\textwidth]{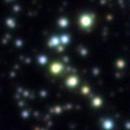}
    \includegraphics[width=0.09\textwidth]{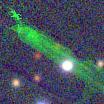}
    \includegraphics[width=0.09\textwidth]{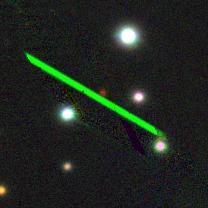}
    \includegraphics[width=0.09\textwidth]{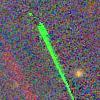}
    % Votes 0.9..1.0
    \includegraphics[width=0.09\textwidth]{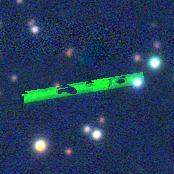}
    \includegraphics[width=0.09\textwidth]{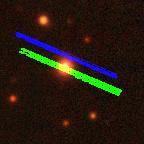}
    \includegraphics[width=0.09\textwidth]{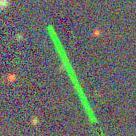}
    \includegraphics[width=0.09\textwidth]{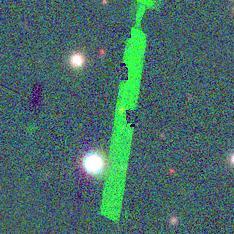}
    \includegraphics[width=0.09\textwidth]{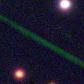}
\caption{
Examples of false positive detections arranged by the neural network score similar to Fig.~\ref{fig:1st_run_examples}.
}
\label{fig:1st_run_false}
\end{figure}

There are at least two types of false detections: 
asterisms (groups of stars that occasionally form a line) and image artefacts in a form of bright strait lines. 
It is clear how they `fool' the network.
Indeed, being narrow elongated objects, especially at low-resolution network inputs of $48\times48$, these `objects' resemble edge-on galaxies, and we did not provide such images as negative examples during the training process. 
More examples of false negative detections are shown in Fig.~\ref{fig:1st_run_false}.
Similar to Fig.~\ref{fig:1st_run_examples}, they are also grouped into rows according to their score. 
As the score increases, the appearance of false detections becomes more and more `galaxy-like'.

To separate false detections from real edge-on galaxies, we performed a visual inspection of all candidates by 12 expert-astronomers.
Each object was examined at least 3 times.
Experts assessed the proximity of the candidate to the edge-on orientation by voting for one of the options: i) a genuine edge-on galaxy, ii) a highly inclined galaxy, $i\gtrsim80^\circ$, iii) a real non-edge-on galaxy that does not satisfy the previous conditions, iv) image artifact, asterism, bright oversaturated star, light reflection, etc.
It turned out that 3,992 detected objects are not galaxies.
We also found 12,882 objects that are disk galaxies, but their inclination is not high enough to hide the structure of their disks, and therefore they can not be considered as edge-on galaxies.
The relatively high percent of galaxies that deviate from desired edge-on orientation can be explained by two facts. 
First, the network input is low-resolution, so when the image is scaled, the minor axis shrinks to just a few pixels, and thus the fine structure of the disk becomes inaccessible to the network.
Second, the EGIS catalog contains a certain percentage of galaxies with inclinations well below 85 degrees, and the network has learned to search for such galaxies.

\subsection{Improvements and the second run}
\label{sec:Run2}

Our pipeline resulted in the largest sample of edge-on galaxies to date~\citep{Makarov2022}, but there is still room for improvement.

First, we have faced a large number of false detections of objects that are not galaxies at all.
Since most of these false positives form groups with a characteristic shape (see fig.~\ref{fig:1st_run_false}), if they were presented in the training sample as negative examples, the network would learn to avoid them.

Second, the contamination of the training sample with galaxies that differ from the edge-on orientation led to a large number of non-edge-on galaxies in the final sample.
A more accurate and ideally larger sample can provide better search quality.

Third, the low resolution of the network input can also affects its effectiveness by smearing out the fine details of the image.

Finally, automatic photometry performed with \textsc{SExtractor}~\citep{Bertin1996} showed a clear shortage of low surface brightness galaxies, especially among ones with small angular sizes. 
We found that, as a rule, these galaxies are lost even before applying the neural network at the stage of the source extraction using the \textsc{detect\_sources} function of \textsc{photutils} package.
Due to the low surface brightness, such galaxies do not meet the specified detection criteria (see section~\ref{sec:search}).

To fix these issue, we have applied the following improvements to our pipeline.
\begin{itemize}
\item We included all non-galactic false detections as negative examples in the new training sample. 
After this step, the training sample became more representative regarding objects that resemble edge-on galaxies, but in fact are not.

\item We extended a list of positive examples of our training sample by 3,482 objects detected during the first run and marked as genuine edge-on galaxy during the visual inspection by at least two experts.

\item The resolution of the network input has been increased to $64\times64$ pixels to capture finer image details.

\item To improve the object selection by the \textsc{detect\_sources} function, we apply it to a combined $g+r+i$ image instead of a single $r$-band image. 
This step significantly increases the detection rate of low surface brightness objects due to a better signal-to-noise ratio of the combined image.

\item Finally, we increased the ensemble size from 5 to 11 models, because we found that the three-out-of-five strategy still occasionally results in misclassification, when three networks erroneously give low or high scores. 
The new six-out-of-eleven approach should be more stable, as there is much less chance of six networks producing a false positive.
\end{itemize}

We ran our modified pipeline again for the same data as described in Section~\ref{sec:search}.

The second run resulted in a significantly increased number of found edge-on galaxies candidates: 132,528 objects have come through new detection procedure on stacked images and received positive votes of six out of eleven networks. 
The visual inspection showed that among these detections there are only 4,348 false positive detections of non-galactic objects.
The full list of candidates is available on the project page~\footnote{\url{https://www.sao.ru/edgeon/catalogs.php?cat=PS1cand2}} in the Edge-on Galaxy Database~\citep{2021AstBu..76..218M}.

\section{The search completeness}

It is obvious that any method can not find all galaxies in the survey images. 
Therefore, it is important to evaluate the completeness of the search.
We assume that the two main parameters of a galaxy that can affect the probability of detection are the central surface brightness of the galaxy and the exponential scale of the light distribution.
When a galaxy is too faint and/or too small in angular size, the number of pixels above the $4\sigma$ detection level may fall below the 15 pixel threshold, and the total size of the extracted object may fall below 48 pixel limit.
Thus, such galaxy will be discarded at the preliminary stage of the \textsc{detect\_objects} function (see section~\ref{sec:search}) and, as a result, it will not be fed to the neural network for classification.
And even if a galaxy is detected by the image segmentation algorithm as a separate object, the neural network may misclassify it due to the low signal-to-noise ratio.

The exact detection level is rather difficult to estimate analytically, because the level of background noise varies significantly between different skycells.
The same galaxy can be easily detected in an image with low background noise, but may be completely missed in some fields with strong background variations due to nearby bright stars, galactic nebulae, poor atmospheric conditions, or in crowded fields such as near the plane of the Milky Way.
Even if a galaxy is big and bright enough to be selected, the detection rate will not be 100\%, because it can overlap with bright oversaturated stars or image artifacts, such as satellite traces, clipping artefacts, etc.

\subsection{Artificial galaxy tests}
\label{sec:completeness}

To estimate the completeness of detection of the edge-on galaxies in the Pan-STARRS1 fields by our method, we performed Monte-Carlo simulations. 
For this purpose, we chose 1500 random skycells, evenly distributed throughout the survey. 
To find the detection probability, we put an image of a real galaxy at a random location on each field, and then perform a full cycle of the edge-on galaxy search, including the objects detection and the neural classification. 
This gives us the fraction of fields in which the galaxy was detected and correctly classified as the edge-on galaxy. 
To find out the dependence of the completeness on the galaxy properties, we scale its size and brightness before placing it in the field.
This allows us to estimate the completeness as a function of the galaxy exponential scale and the central surface brightness. 
To eliminate the possible influence of the features of an individual object, we run this simulation for 20 random galaxies and average the results.

\begin{figure}
\centering
\includegraphics[width=\columnwidth]{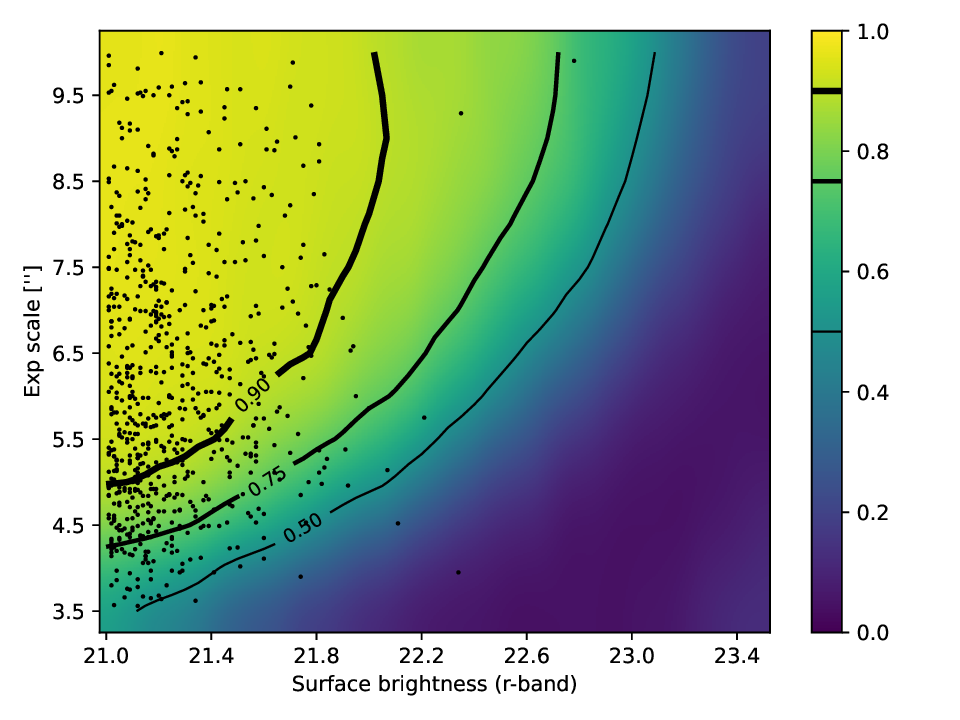}
\caption{
Fraction of detected galaxies as a function of the disk exponential scale and the central surface brightness. 
Black dots are galaxies from the EGIS catalog~\citep{Bizyaev2014}.
}
  \label{fig:completeness}
\end{figure}

Figure~\ref{fig:completeness} shows the results of the test for a grid of central surface brightness ranging from 21 to 23.5~mag/sq.arcsec in the $r$-band and for an exponential scale ranging from 3 to 10$^{\prime\prime}$. 
One can see that, as expected, the completeness is higher for large and bright galaxies and reaches almost 100\% for them, while for small and faint ones it decreases. We expect 90\% completeness for galaxies having the central surface brightness less than $\approx22$~mag/sq.arcsec in the $r$-band and the disk scale bigger than $\approx5^{\prime\prime}$.
For comparison, we place EGIS-galaxies~\citep{Bizyaev2014} on this diagram, which indicates that our sample should contain more faint galaxies compared to this catalog.

\subsection{Comparison with RFGC \& EGIS}

We test the efficiency of the method in finding real galaxies from the RFGC~\citep{Karachentsev1999} and EGIS~\citep{Bizyaev2014} catalogs.
Both catalogs were created as a result of visual inspection of galaxy images.
The main difference is the selection criteria.
In the older RFGC-catalog, the galaxies were selected by their axis ratio, $a/b\ge7$, measured on photographic prints of the POSS-I and ESO/SERC surveys.
While in the EGIS-catalog the edge-on galaxies were chosen by visual inspection of a preselected sample of objects automatically found in the SDSS survey, with $g$-band axes ratio $a/b>3$ \citep[for details see][]{Bizyaev2014}.
As a result, the RFGC-catalog contains predominantly late-type bulgeless galaxies.
On the other hand, the type distribution in the EGIS-catalog is more uniform, but due to automatic selection, some known edge-on galaxies were lost and had to be added manually.

\begin{figure*}
\centering
\begin{tabular}{cc}
\includegraphics[width=\columnwidth]{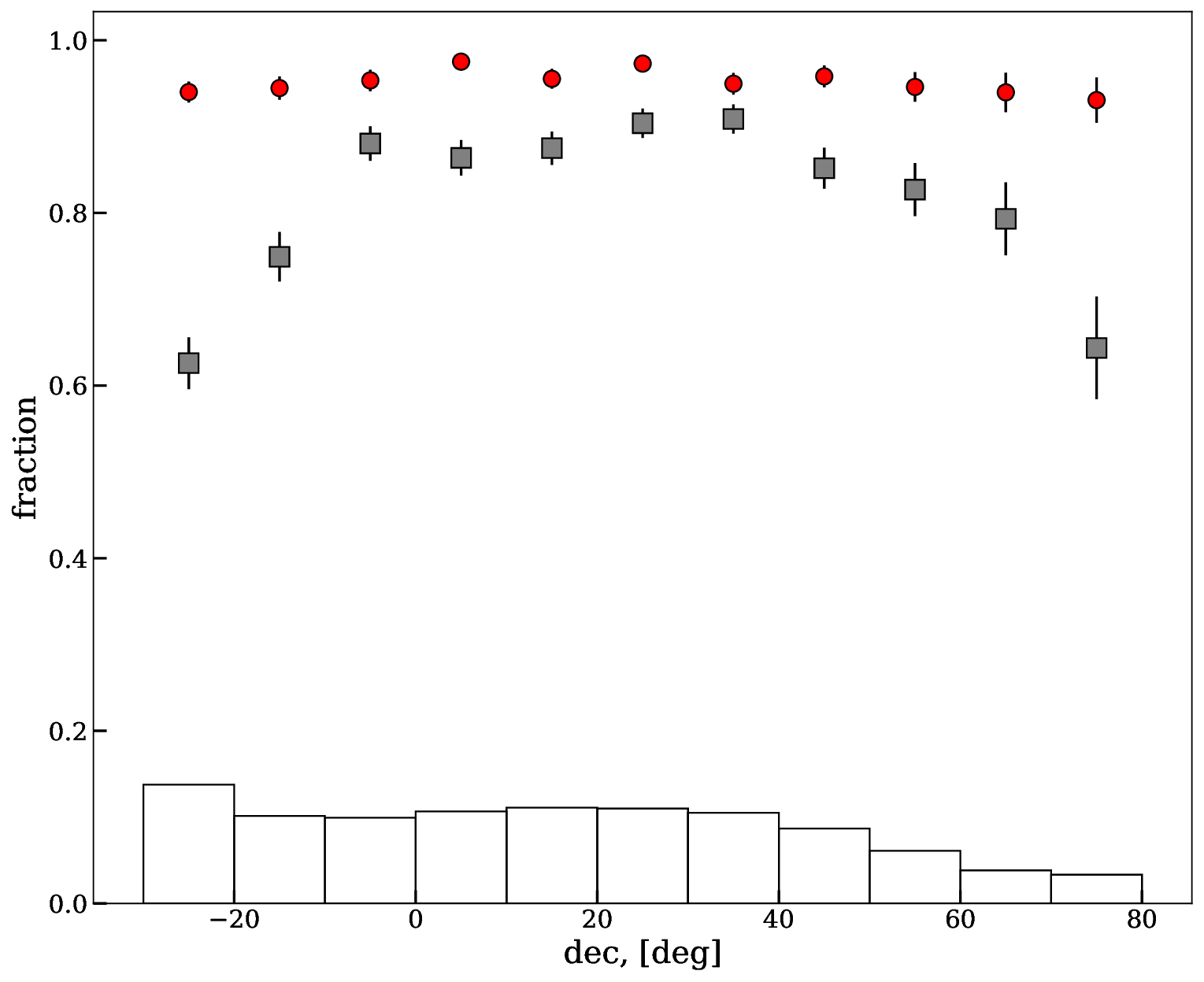} &
\includegraphics[width=\columnwidth]{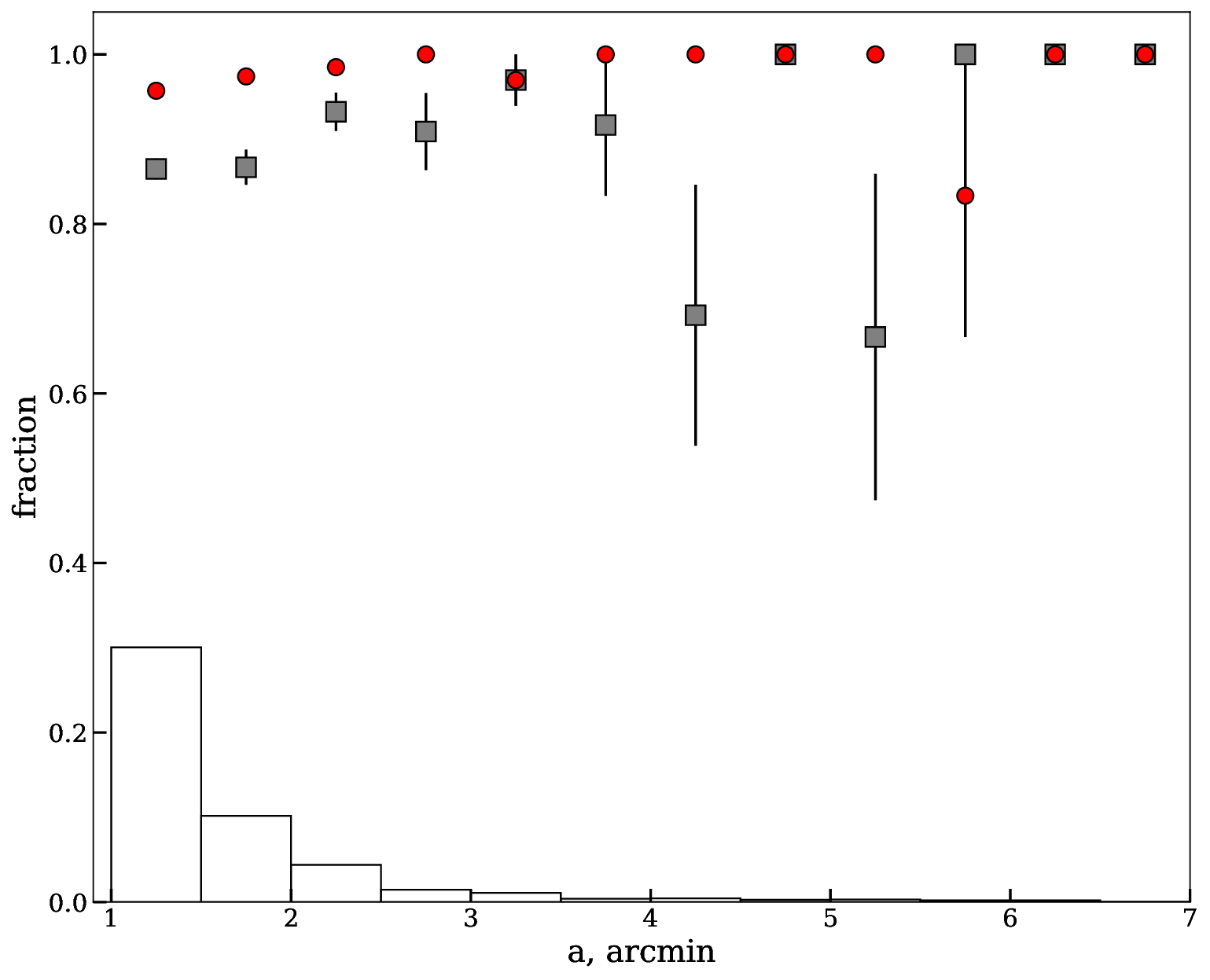} \\
\includegraphics[width=\columnwidth]{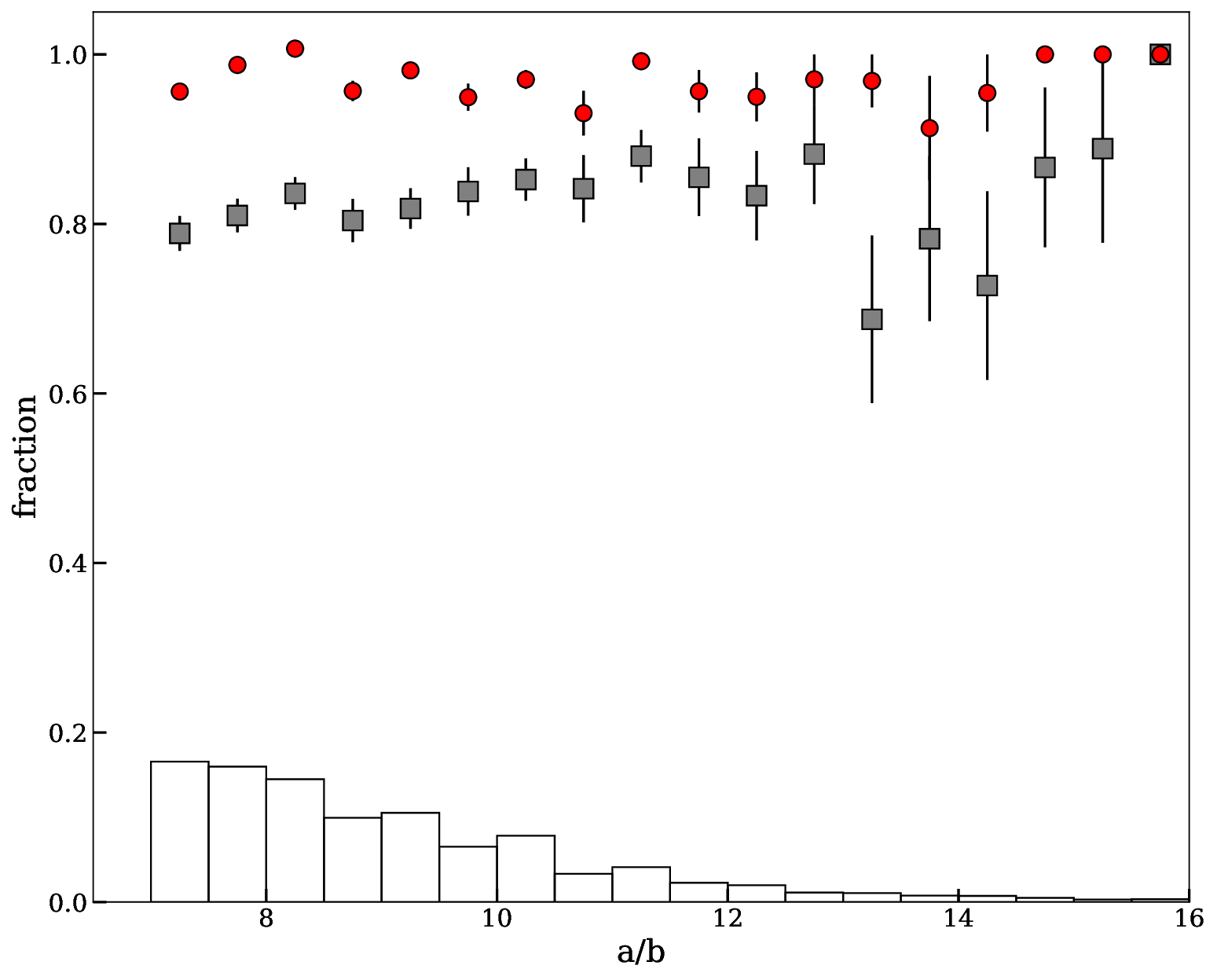} &
\includegraphics[width=\columnwidth]{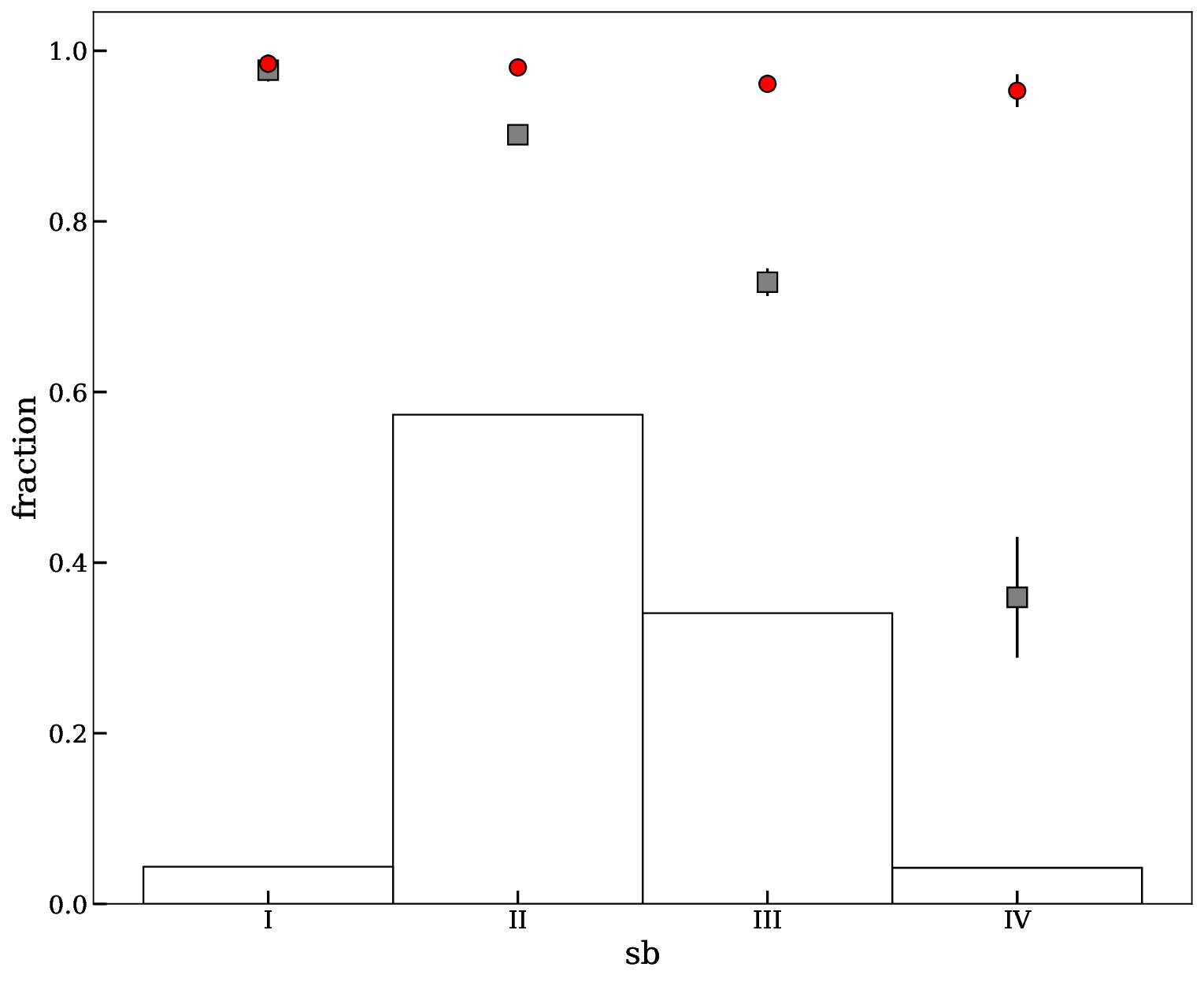}
\end{tabular}
\caption{
The search completeness for the RFGC~galaxies~\citep{Karachentsev1999}.
The gray squares correspond to the first run.
The red circles show the detection completeness in the second run.
The bars correspond to the 1-sigma confidence interval (Wald interval) of the binomial proportion~\citep{SeanWallis}.
The histogram shows the distribution of RFGC galaxies.
The top left panel is for the declination.
The top right is for the major axis in arcmin.
The bottom left is for the axes ratio.
The bottom right is for the surface brightness class.
}
\label{fig:rfgc}
\end{figure*}

In the case of the RFGC catalog, we estimated completeness only for galaxies with $\mathrm{Dec.}>-30^\circ$ in the region of intersection with the Pan-STARRS1 survey.
In the second pass, we were able not only to significantly improve the level of object detection from 82 to 97\%, but also the detection quality.
Figure~\ref{fig:rfgc} shows that the search completeness over the sky became much more uniform as we approached the pole and the southern edge of the Pan-STARRS1 survey (the top left panel).

The photometric depth of the Pan-STARRS1 survey~\citep[][see fig.~17]{PS1Survey} drops slightly near the circumpolar region, as well as at low latitudes. 
This reflects in a decrease of the detection completeness in the first run, which was carried out only in one $r$-filter. 
The summation of the three $g, r, i$ frames that was added in the second run increased the signal-to-noise ratio of faint galaxies and solved this problem. 
The completeness of the second run remains almost constant regardless of latitude.

Also there is no significant trends in the detection level with the object size (the top right panel) and the axis ratio (the bottom left panel).
The greatest progress has been made in the detection of low surface brightness galaxies.
All galaxies in the RFGC catalog are divided into 4 classes of surface brightness from the brightest (I) to the dimmest (IV).
As it can be seen from the bottom right panel, in the first version of the algorithm, we were catastrophically short of low surface brightness galaxies.
On the second pass, the drop of the detection rate with surface brightness is very modest.

\begin{figure*}
\centering
\begin{tabular}{cc}
\includegraphics[width=\columnwidth]{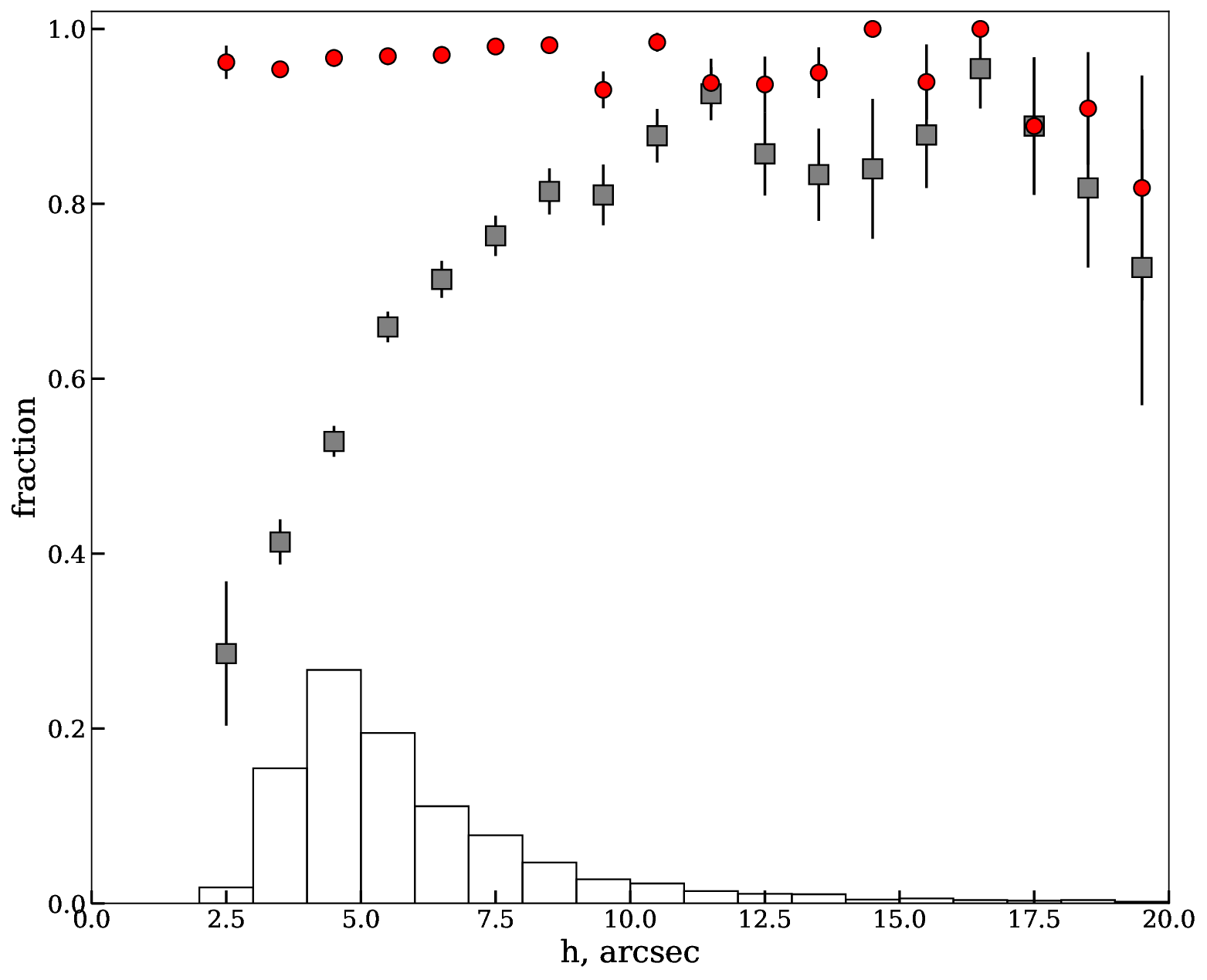} &
\includegraphics[width=\columnwidth]{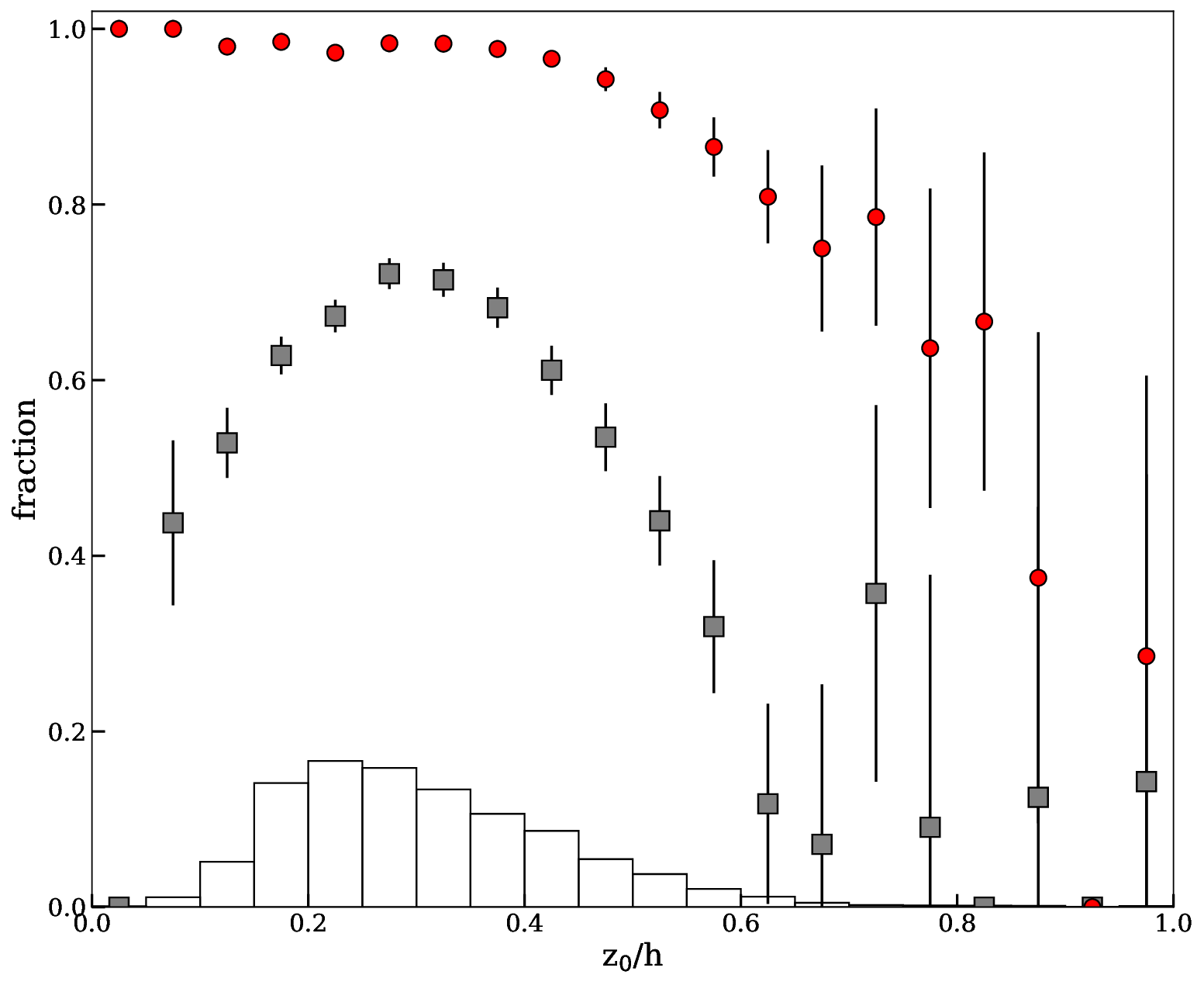} \\
\includegraphics[width=\columnwidth]{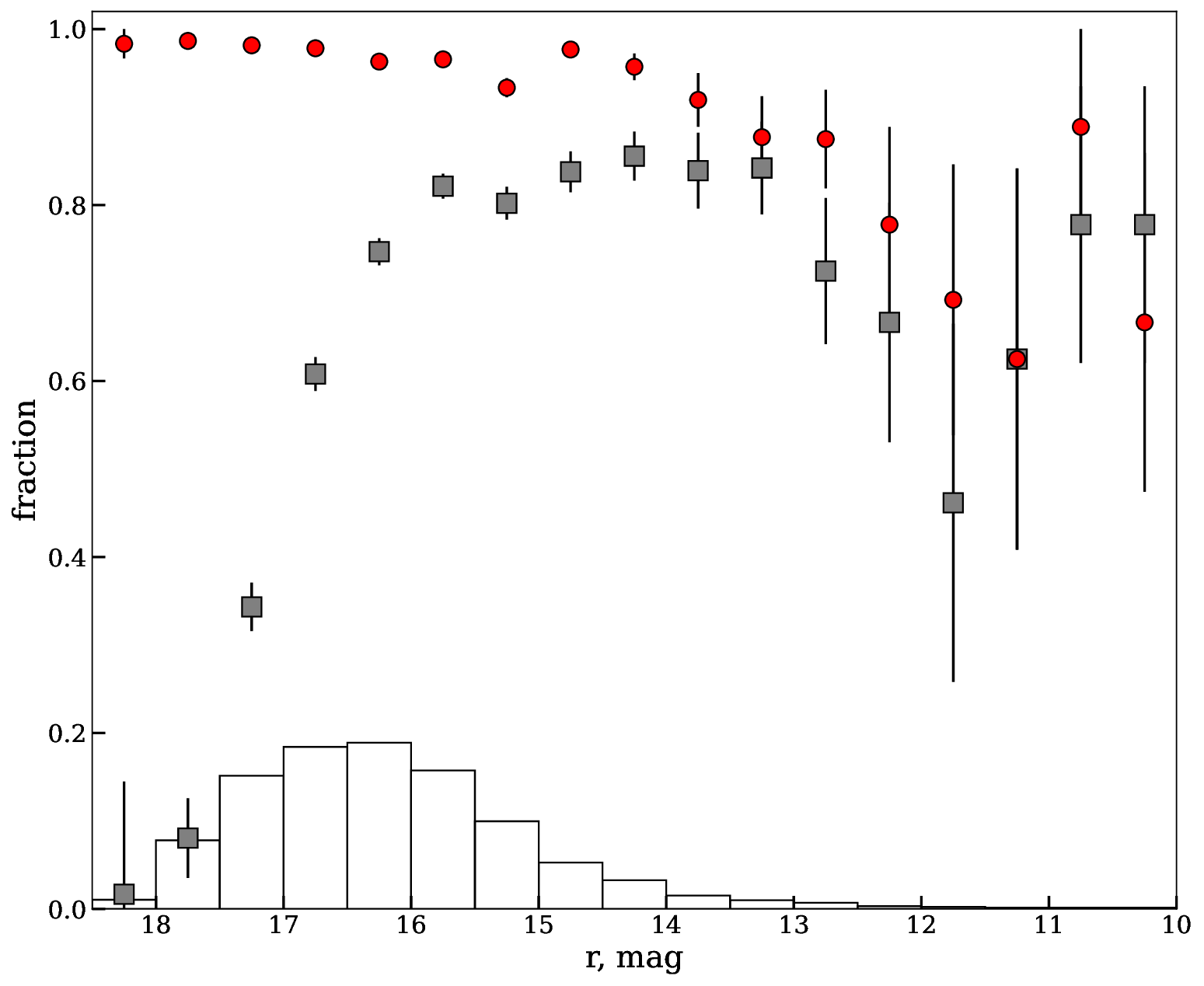} &
\includegraphics[width=\columnwidth]{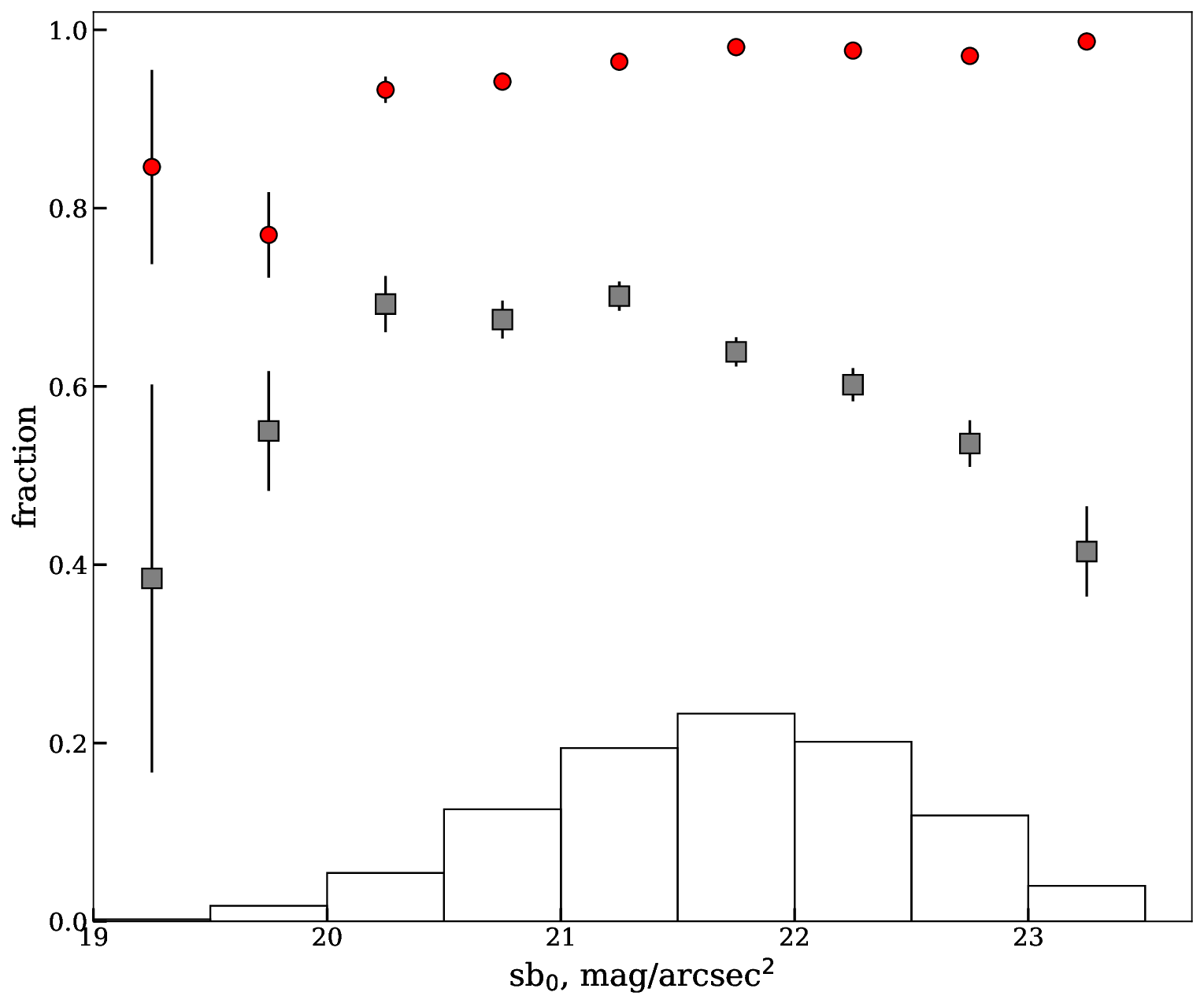}
\end{tabular}
\caption{
The search completeness for the EGIS~galaxies~\citep{Bizyaev2014}.
The designations are similar to fig.~\ref{fig:rfgc}.
The top left panel is for the exponential scale of the disk of EGIS-galaxies.
The top right is for the vertical-to-radial scale ratio of the disk.
The bottom left is for the total magnitude of the galaxy.
The bottom right is for the central surface brightness.
}
\label{fig:egis}
\end{figure*}

The area of the SDSS survey lies entirely within the Pan-STARRS1 survey.
Therefore, all EGIS-galaxies should have been passed through our algorithm.
For all EGIS-galaxies, aperture photometry, structural bulge-disk decomposition, and parameters of the exponential disk (central brightness, vertical and radial scales) were determined in the original work by~\citet{Bizyaev2014}.
This allows us to make more direct estimates of the completeness depending on galaxy disk parameters.
Figure~\ref{fig:egis} also shows huge improvements in the second run compared to the first one.
In the first run, the detection of the smallest and faintest galaxies was extremely poor, falling below 50\% for galaxies with the radial disk scale less than 5~arcsec (the top left panel) or fainter than 17~mag in the $r$-band (the bottom left panel).
The second run solved this issue.
It is necessary to note a smooth decline of the completeness for the thickest galaxies $z_0/h\gtrsim0.4$ (the top right panel).
This happens for two reasons.
The inspection shows that the some thick EGIS-galaxies are not real edge-on galaxies and do not pass the classification stage.
The second reason is that the real thick edge-on galaxy is a quite rare object. 
Since its axis ratio is similar to that of a normal galaxy inclined less than $60^\circ$, this requires special attention during the training process.
The completeness is 98\% if the thickest galaxies $z_0/h\gtrsim0.4$ are excluded from the consideration (96\% for the entire sample).
One more point that should be noted is the loss of the brightest (the bottom left) and largest (the top left) galaxies with high central surface brightness (the bottom right).
Our methodology failed to select 23 out of 101 galaxies with $r>13$~mag. The main reason of this is an upper limit that we imposed on the candidate size on the stage of object selection with \textsc{photutils}. Adjacent Pan-STARRS1 fields have a 2 arcminutes overlapping regions at their borders, so if a galaxy image is less that 2 arcminutes in size, it is guaranteed to fit entirely inside one or another Pan-STARRS1 field. Images of larger galaxies, on the other hand, can be split between two or more adjacent fields depending on their coordinates and orientation and to classify them properly one has to concatenate these fields first. This would require a considerable amount of an additional computational payload, whereas all such large galaxies are surely already included in existing catalogues of edge-on galaxies.
%It also turned out that the vast majority of the brightest lost galaxies (19 out of 23) are also extremely thick $z_0/h>0.4$, moreover, six of them have $z_0>h$.
%Thus, the loss of bright galaxies is a particular case of the problem selecting thick edge-on galaxies.
%In addition, images of these bright and large galaxies show a detailed complex internal structure, which can be interpreted by the ANN as an indication that they are not seen edge-on.

\section{Conclusions}

In this work, we describe an algorithm for searching for a specific class of objects, the so-called edge-on galaxies, in the Pan-STARRS1 survey images using ANNs.
This solution is an alternative to the classical approach based on visual inspection of images or selection by axes ratio from automatically generated object catalogs.

The final version of our pipeline consists of two main parts: 
the selection of objects in an image using the \textsc{python} photometry package \textsc{photutils}~\citep{2020zndo...4044744B}
and the objects classification by an ensemble of ANN models trained to find edge-on galaxies.

The search for objects is carried out in a combined image from three images in the $g$, $r$ and $i$ bands of the Pan-STARRS1 survey. 
An object is considered detected if its major axis exceeds 48 pixels with at least 15 connected pixels are above $4\sigma$ of background noise.
After that, we extract the object images in the $g$, $r$, $i$ bands, scale them to $64\times64$ pixels, and combine them into a 3D array for ANN classification. 

The resulting stack of $64\times64\times3$ pixels is fed to an ensemble of eleven ANN models.
Each model consists of three rounds of convolutional blocks.
Each block, in turn, consists of two convolutional layers, followed by batch normalization, max-pooling and dropout layers.
The network architecture is completed with two fully connected layers with 500 and 2 neurons, which perform image classification and evaluate the classification of the object as the edge-on galaxy (see Section~\ref{sec:networkArchitecture} for details).

%Each network was independently trained on a sample of 3,482 true edge-on galaxies.
%This sample was obtained by the visual inspection by experts of 26,719 candidates received during the first run of the network (each candidate was checked by at least three experts, and only those who received at least two votes were marked as true edge-on galaxies).
Each network was independently trained on a sample of 9,229 positive examples.
This sample was formed by combining 5,747 edge-on galaxies from the EGIS catalog~\citep{Bizyaev2014} and additional set of 3,482 true edge-on galaxies selected during the visual inspection by experts of 26,719 candidates received during the first run of the network (at least two experts had to mark these objects as edge-on galaxies).

As negative examples, we used a sample of real non-edge-on galaxies, supplemented by empty fields, images of stars, and, which turned out to be extremely important, examples of image defects and various configurations that mimic real edge-on galaxies.

All these steps allowed us to construct effective and reliable method for searching for the edge-on galaxies.
We found 128,180 real candidates in edge-on galaxies with only 4,348 wrong detections (image artifacts, asterisms etc.).
The artificial galaxy test shows that the 90\% completeness level is achieved for edge-on galaxies with the central surface brightness of 22~mag$/\square^{\prime\prime}$ and the disk radial scale of $5^{\prime\prime}$.
Comparison with known catalogs of edge-on galaxies demonstrates 97\% detection rate in a wide range of parameters.
There are no significant trends of the completeness with position in the sky, object size and total magnitude, as well as its surface brightness.
The method works well for the disk galaxies thinner than $z_0/h=0.4$, where $z_0$ and $h$ denote the vertical and radial scales, respectively, of an isothermal exponential stellar disk. 
Applying this method to the Pan-STARRS1 survey~\citep{PS1Survey} allowed us to obtain the most complete and largest sample of galaxies to date.
Data access is provided by the Edge-on Galaxy Database~\citep{2021AstBu..76..218M} on the web-page of the project~\footnote{\url{https://www.sao.ru/edgeon/catalogs.php?cat=PS1cand2}}.
For each candidate, the score statistics of the proximity to the edge-on orientation, obtained by the ANN, is given, as well as set of its photometric parameters (coordinates of the center, the galaxy size along the major and minor semiaxes, aperture magnitude).

\section*{Acknowledgements}
This research was supported by the Russian Science Foundation grant \textnumero~19--12--00145.

The authors thank the anonymous referees for their helpful comments that improved the quality of the manuscript.

The Pan-STARRS1 Surveys (PS1) and the PS1 public science archive have been made possible through contributions by 
the Institute for Astronomy, the University of Hawaii, the Pan-STARRS Project Office, the Max-Planck Society and its participating institutes, 
the Max Planck Institute for Astronomy, Heidelberg and the Max Planck Institute for Extraterrestrial Physics, Garching, 
The Johns Hopkins University, Durham University, the University of Edinburgh, the Queen's University Belfast, 
the Harvard-Smithsonian Center for Astrophysics, the Las Cumbres Observatory Global Telescope Network Incorporated, 
the National Central University of Taiwan, the Space Telescope Science Institute, 
the National Aeronautics and Space Administration under Grant No. NNX08AR22G 
issued through the Planetary Science Division of the NASA Science Mission Directorate, 
the National Science Foundation Grant No. AST-1238877, 
the University of Maryland, Eotvos Lorand University (ELTE), the Los Alamos National Laboratory, and the Gordon and Betty Moore Foundation.

We acknowledge the usage of the HyperLeda\footnote{\url{http://leda.univ-lyon1.fr/}} database~\citep{Makarov2014}.

%\appendix

%\section{Appendix title 1}

%\section{Appendix title 2}

\bibliographystyle{elsarticle-harv} 
\bibliography{bibliography}



%% For example:

%% \bibitem[Aladro et al.(2015)]{Aladro15} Aladro, R., Martín, S., Riquelme, D., et al. 2015, \aas, 579, A101

%%\end{thebibliography}

\end{document}